\documentstyle[aps,psfig,floats,epsf,twocolumn,prb]{revtex}

\begin{document}
\tighten
\draft
\twocolumn[
\hsize\textwidth\columnwidth\hsize\csname @twocolumnfalse\endcsname

\title{Numerical Renormalization Group for Quantum Impurities in a
Bosonic Bath}
\author{Ralf Bulla$^*$, Hyun-Jung Lee$^*$, 
Ning-Hua Tong$^*$, and Matthias Vojta$^\dagger$}
\address{$^*$Theoretische Physik III, Elektronische Korrelationen und
Magnetismus, Institut f\"ur Physik, Universit\"at Augsburg,  D-86135 Augsburg,
Germany}
\address{$^\dagger$Institut f\"ur Theorie der Kondensierten Materie,
Universit\"at Karlsruhe,
D-76128 Karlsruhe,
Germany}
\date{\today}
\maketitle

\widetext
\begin{abstract}
We present a detailed description of the recently proposed
numerical renormalization group method for models of quantum impurities
coupled to a bosonic bath.
Specifically, the method is applied to the spin-boson model, both in
the Ohmic and sub-Ohmic cases.
We present various results for static as well as dynamic quantities
and discuss details of the numerical implementation, e.g., the discretization
of a bosonic bath with arbitrary continuous spectral density,
the suitable choice of a finite basis in the bosonic Hilbert space,
and questions of convergence w.r.t. truncation parameters.
The method is shown to provide high-accuracy data
over the whole range of model parameters and temperatures,
which are in agreement with exact results and other numerical data
from the literature.
\end{abstract}
\pacs{PACS: 05.10.Cc (Renormalization Group methods),
05.30.Jp (Boson systems)}
\vspace*{0.7cm}
]

\narrowtext

\section{Introduction}
The Numerical Renormalization Group (NRG) is known as a powerful tool
for the investigation of quantum impurity problems,
where a quantum system with a finite number of internal degrees of freedom
(the impurity) couples to an infinite system of non-interacting fermions
with a continuous density of states (the bath)
\cite{Wil75,Kri80,Hewson,Costi99a,RBAdvPhys}.
The NRG combines numerically exact diagonalization with the
idea of the renormalization group, where progressively smaller energy scales
are treated in the course of the calculation.
NRG calculations are non-perturbative and -- thanks to the logarithmic
energy discretization -- are able to access arbitrarily small
energies and temperatures.
Besides providing thermodynamic quantities like susceptibility, entropy, and
magnetization, the NRG can be used to compute dynamic observables directly on the real
frequency axis.

While the NRG was originally developed by Wilson \cite{Wil75} for the Kondo model,
it was later applied to a variety of more complex impurity models
with one or more fermionic baths, being able to handle, e.g., two-channel
and multi-impurity physics \cite{multinrg,twochnrg}.
As a recent extension, impurity models with a fermionic bath and
a single bosonic mode have been treated, with the so-called Anderson-Holstein impurity
model being the paradigmatic example \cite{holsteinnrg}.
Interesting applications of the NRG include its use
within dynamical mean-field theory (DMFT)\cite{MV,Geo96}.
There, the electronic self-energy of a lattice model of correlated electrons
is approximated by a local function in space,
and the lattice model is mapped onto a single-impurity model supplemented
by a self-consistency condition.
Using DMFT-NRG, the Mott transition of the Hubbard model has been investigated in
detail, both at zero and finite temperatures \cite{dmftnrg,BCV}.

The objective of this paper is an important extension of the NRG method,
namely the application to quantum impurities coupled to a bosonic bath
with a continuous spectral density (in contrast to the single boson mode in
Ref.~\onlinecite{holsteinnrg}).
We have recently given a short account on this development\cite{BTV};
the purpose here is a detailed description of this novel NRG
application.
To be specific, most of our presentation will focus on the spin-boson model,
with the Hamiltonian
\begin{equation}
H=-\frac{\Delta}{2}\sigma_{x}+\frac{\epsilon}{2}\sigma_{z}+
\sum_{i} \omega_{i}
     a_{i}^{\dagger} a_{i}
+\frac{\sigma_{z}}{2} \sum_{i}
    \lambda_{i}( a_{i} + a_{i}^{\dagger} ) \ .
\label{eq:sbm}
\end{equation}
This model naturally arises in the description of quantum
dissipative systems\cite{Leggett,Weiss}:
The dynamics of the two-state system, represented by the Pauli matrices $\sigma_{x,z}$,
is governed by the competition between the tunneling term $\Delta$ and the friction
term $\lambda_{i}(a_{i}+a_{i}^{\dagger})$.
The $a_i$ constitute a bath of harmonic oscillators responsible for the damping,
characterized by the bath spectral function
\begin{equation}
    J\left( \omega \right)=\pi \sum_{i}
\lambda_{i}^{2} \delta\left( \omega -\omega_{i} \right) \,.
\end{equation}
Clearly, most interesting are gapless spectra, $J(\omega)>0$ for $0<\omega<\omega_c$,
with $\omega_c$ being a cutoff energy.
In the infrared limit, the energy dependence of $J(\omega)$ for $\omega\to 0$
determines the system's behavior, where power-laws are of particular importance.
Discarding high-energy details of the spectrum, the standard parametrization is
\begin{equation}
  J(\omega) = 2\pi\, \alpha\, \omega_c^{1-s} \, \omega^s\,,~ 0<\omega<\omega_c\,,\ \ \ s>-1 \,.
\label{power}
\end{equation}
The case $s=1$ is known as Ohmic dissipation \cite{Leggett}, where the
spin-boson model has a delocalized and a localized zero-temperature phase,
separated by a Kosterlitz-Thouless transition (for the unbiased case of $\epsilon=0$).
In the delocalized phase, realized at small dissipation strength $\alpha$,
the ground state is non-degenerate and represents a (damped) tunneling
particle.
For large $\alpha$, the dissipation leads to a localization of the particle in
one of the two $\sigma_z$ eigenstates, thus the ground state is doubly
degenerate.

Bath spectra with exponents $s>1$ ($s<1$) are called super-Ohmic (sub-Ohmic):
In the super-Ohmic case, the system is always delocalized with weak
damping; the sub-Ohmic case is more involved and will be discussed below.
Besides simple power-law spectra, more complicated bath properties
can arise in a number of situations, e.g., structured baths, consisting of
an Ohmic part and modes sharply peaked at certain energies, have been
considered recently \cite{structbath,centralspin}.

The spin-boson model has found applications in a wide variety of physical
situations\cite{Leggett,Weiss}:
mechanical friction, damping in electric circuits,
decoherence of quantum oscillations in qubits\cite{Cos03,dima,wilhelm},
impurity moments coupled to bulk magnetic fluctuations\cite{bosekondo}, and
electron transfer in biological molecules\cite{Garg,MuehlbacherEgger}.

Considering this wealth of applications, numerical methods to reliably
deal with the spin-boson and related models for all temperatures are
highly desirable.
In the past, quantum Monte-Carlo simulations\cite{qmcsb}
have been used, which, however,
cannot work at arbitrarily low temperature, and cannot easily extract
dynamical information on the real frequency axis.
Density-matrix renormalization techniques, as employed in 
Ref.~\onlinecite{nishiyama}, 
circumvent this problem, but are not able to resolve
very small energy scales.
In Ref.~\onlinecite{Costi98}, the NRG with a {\em fermionic} bath
has been used, exploiting the well-established mapping of the
Ohmic spin-boson model to an anisotropic Kondo model.
Such a mapping is only valid for frequencies $\omega\ll \omega_c$
(which nevertheless encompasses most of the interesting physics),
and, more seriously, is restricted to the Ohmic case \cite{MVLF}.

In a recent paper, we have presented a formulation of the NRG
directly for a bosonic bath, and applied it to the
spin-boson model\cite{BTV}.
While we could accurately reproduce known results for the Ohmic case,
we also found that the sub-Ohmic model displays two phases as well
(in agreement with Refs.~\onlinecite{KM,spohn})
which are separated by a non-trivial quantum phase transition.
Remarkably, this phase transition was not systematically
investigated before. 
We have studied the properties of the corresponding quantum
critical points --
in the phase diagram (see Fig.~1 of Ref.~\onlinecite{BTV})
those form a line, parametrized by the bath exponent $s$,
which terminates in the Kosterlitz-Thouless transition at $s\!=\!1$.
Near $s\!=\!1$ we could make contact with analytical renormalization
group results, originally formulated by Kosterlitz in the context of an
Ising model with long-range $1/r^{1+s}$ interaction\cite{koster}.

The purpose of this paper is
(i)   to present in detail the implementation
of the bosonic NRG method for the spin-boson model,
(ii)  to discuss various strategies to set up the iteration scheme for the bosonic NRG,
(iii) to demonstrate its feasibility by studying, in particular,
the case of Ohmic damping; and compare our data with a variety
of results from the literature, and
(iv)  to discuss possible future applications of the bosonic NRG.
The physics of the sub-Ohmic spin-boson model is very rich
due to the presence of a line of boundary quantum critical points;
a full account of the universal critical behavior, studied using analytical
and numerical methods, will be given in a forthcoming 
publication\cite{VTB}.

The remainder of the paper is organized as follows:
In Sec.~\ref{sec:bosonicNRG} we introduce the formulation of the
NRG for bosonic systems and highlight important differences which
occur compared to the fermionic NRG.
In particular, the choice of appropriate bosonic basis states,
which are required to accurately describe certain strong-coupling fixed
points, is discussed, with details given in Appendix A and B.
Section \ref{sec:flow} analyzes the NRG flow and the low-energy
fixed points, the phase boundaries and issues of numerical convergence
as function of the discretization parameters.
In Sec.~\ref{sec:thdyn} we turn to thermodynamic observables calculated
using the bosonic NRG, such as entropy and specific heat, together
with their scaling behavior.
Finally, Sec.~\ref{sec:dyn} is devoted to dynamical quantities,
where we focus on the symmetrized impurity spin autocorrelation function.
We close with a summary and discussion of applications and extensions
of the bosonic NRG.

\section{The bosonic NRG}
\label{sec:bosonicNRG}
The bosonic NRG can be applied to a wide range of
quantum impurity problems involving a
bosonic bath with a continuous (and in particular
gapless) spectrum. The following discussion
of the technical details of this method concentrates
on the spin-boson model, the first application
of the bosonic NRG \cite{BTV}. The general concepts
are valid for the study of other bosonic impurity
models as well.

The purpose of this section is twofold: we
want to discuss in detail the technical steps
of the calculations in Ref.~\onlinecite{BTV}. In addition,
we introduce an alternative strategy to set up
the NRG procedure, the so-called {\em star}-NRG, in contrast
to the {\em chain}-NRG used in Ref.~\onlinecite{BTV}.
The use of the star-NRG is related to the choice of an
optimized set of basis states.
As will be discussed in Secs.~\ref{sec:genstrat} and \ref{sec:3D},
the star-NRG allows for an efficient construction of the NRG basis
which solves the problem of the boson number divergence
occurring in the localized phase for sub-Ohmic damping.

Let us start with a form of the spin-boson model which
is most convenient for the NRG procedure:
\begin{equation}
 H= H_{\rm loc} +
  \int\limits_{0}^{1} {\rm d}\varepsilon \,
g(\varepsilon) a_{\varepsilon}^{\dagger} a_{\varepsilon}
 + \frac{\sigma_{z}}{2}
\int\limits_{0}^{1} {\rm d}\varepsilon \, h(\varepsilon )
   (a_{\varepsilon} + a_{\varepsilon}^{\dagger} )
\label{eq:Hsbm-int}
\end{equation}
with $H_{\rm loc} = -\Delta\sigma_{x}/2 + \epsilon\sigma_{z} / 2$.
In this model, $g(\varepsilon)$ characterizes the dispersion
of a bosonic bath in a one-dimensional representation,
with upper cutoff 1 for $\varepsilon$. The coupling
between the spin and the bosonic bath is given by $h(\varepsilon)$.
These two energy-dependent functions are related to the spectral
function $J(\omega)$ via
\begin{equation}
  \frac{1}{\pi}J(x)=\frac{d \varepsilon(x)}{d x} h^{2}
\left[ \varepsilon(x) \right]
  \quad \left( x \in \left[0, \omega_{c} \right] \right) \, ,
\label{eq:h}
\end{equation}
where $\varepsilon(x)$ is the inverse function of $g(x)$,
$g[\varepsilon(x)]=x$.
For a given $J(x)$, eq.~(\ref{eq:h}) does {\em not} determine both $g$ and
$h$ independently. Therefore, as shown below in eq.~(\ref{eq:hdisc}),
a specific choice of $h$ is used to simplify the calculations.

\begin{figure}[!t]
\epsfxsize=2.0in
\centerline{\epsffile{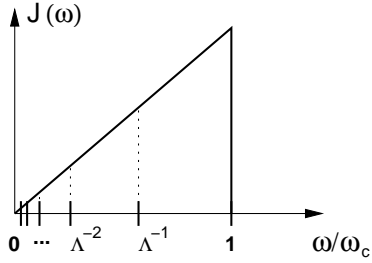}}
\caption{
Logarithmic discretization of the bath spectral function in intervals
 $[\Lambda^{-(n+1)},\Lambda^{-n}]$
($n=0,1,2,\ldots$); typical values of the NRG discretization
parameter $\Lambda$ as used in the bosonic NRG are
$\Lambda=1.5 \ldots 3.0$.
}
\label{fig:Jdisc}
\end{figure}

\subsection{Logarithmic Discretization}

The NRG procedure starts by dividing the interval
$[0,1]$ into intervals $[\Lambda^{-(n+1)},\Lambda^{-n}]$
($n=0,1,2,\ldots$, see Fig.~\ref{fig:Jdisc}).
The width of each interval is
\begin{equation}
   d_n =
                          \Lambda^{-n}(1-\Lambda^{-1})
\end{equation}
Within each interval we introduce a complete set of orthonormal functions:
\begin{equation}
   \psi_{np} (\varepsilon) = \left\{ \begin{array}{ll}
                          \frac{1}{\sqrt{d_n}} e^{ i\omega_n p \varepsilon}
                          & \mbox{for}\ \Lambda^{-(n+1)} < \varepsilon <
                            \Lambda^{-n} \\
                          0 & \mbox{outside this interval}
                                         \end{array}  \right. ,
\end{equation}
($p=0,\pm1,\pm2,\ldots$ and $\omega_n = 2\pi / d_n $).
The operators $a_\varepsilon^{(\dagger)}$ appearing in the Hamiltonian
(\ref{eq:Hsbm-int})
can be represented in
this basis:
\begin{eqnarray}
a_{\varepsilon}&=& \sum \limits_{np} a_{np} \psi_{np}(\varepsilon) \\
 a_{\varepsilon}^{\dagger}&=& \sum\limits_{np} a_{np}^{\dagger}
\psi_{np}^{*}(\varepsilon) \ .
\end{eqnarray}
We then choose the function
$h(\varepsilon)$ to be a constant $h_n$ in each intervall of the
logarithmic discretization:
\begin{equation}
   h(\varepsilon)= h_n=\left[ \frac{1}{\Lambda^{-n}-\Lambda^{-(n+1)}}
    \int_{\Lambda^{-(n+1)} \omega_{c}}^{\Lambda^{-n} \omega_{c}} \frac{1}{\pi}J(\omega) \, {\rm d}\omega \right]^{1/2} \label{eq:hdisc}
\end{equation}
for $\varepsilon \in \left[ \Lambda^{-(n+1)}, \Lambda^{-n} \right]$.
With this choice, the impurity (the spin-operator $\sigma_z$)
couples to the $p\!=\!0$ component of the bosonic operators
$a_{np}$ and $a_{np}^{\dagger}$ only (the same strategy has
been used in the case of a fermionic bath with non-constant
density of states, see Ref.~\onlinecite{BPH}).

The next step is to write the Hamiltonian (\ref{eq:Hsbm-int}) in the basis
$a_{np}$ and $a_{np}^{\dagger}$; the $p\!\ne\!0$ components
of these operators are still present through their coupling
to the $p\!=\!0$ components in the free bath term. The main
approximation of the bosonic NRG at this point is to drop
this coupling, in close analogy to the fermionic case
(see Refs.~\onlinecite{Wil75,Kri80}).
This approximation becomes exact in the limit
$\Lambda\to 1$. Nevertheless, a careful check of its validity
is necessary and will be discussed in Sec.~\ref{sec:3B}.

With the $p\!\ne\!0$ components completely decoupled from
the impurity, we drop the $p\!=\!0$ index in the operators
$a_{np=0}$ and $a_{np=0}^{\dagger}$ and arrive at a Hamiltonian
of the form:
\begin{equation}
H_{\rm s} =
H_{\rm loc} + \sum\limits_{n=0}^{\infty} \xi_{n}a_{n}^{\dagger}a_{n}
  + \frac{\sigma_z}{2\sqrt{\pi}}
  \sum\limits_{n=0}^{\infty} \gamma_{n} \left(a_{n}+a_{n}^{\dagger} \right),
\label{eq:hstar}
\end{equation}
with
\begin{equation}
    \xi_{n}\!= \! \gamma_n^{-2}
    \int_{\Lambda^{-(n+1)}\omega_c}^{\Lambda^{-n}\omega_c}
    \!\!\!\!\!\!\!\!\!\!{\rm d}x \,J(x)x \, ,~~
    \gamma_{n}^2=
    \int_{\Lambda^{-(n+1)}\omega_c}^{\Lambda^{-n}\omega_c}
    \!\!\!\!\!\!\!\!\!\!{\rm d}x \, J(x) .
\end{equation}
The label $H_{\rm s}$ is introduced to distinguish this ``star''-Hamiltonian from
the ``chain''-Hamiltonian $H_{\rm c}$ (see eq.~(\ref{eq:hchain}) below).
The $\xi_{n}$ and the $\gamma_{n}$ can be easily evaluated for a
bath spectral function of the form given in eq.~(\ref{power}):
\begin{eqnarray}
  \xi_{n} & = & \frac{s+1}{s+2}
               \frac{1-\Lambda^{-(s+2)}}{1-\Lambda^{-(s+1)}} \omega_c
   \Lambda^{-n} \nonumber \\
   \gamma^2_{n}  & = & \frac{2\pi\alpha}{s+1}  \omega_c^2
                       \left( 1-\Lambda^{-(s+1)} \right)
                       \Lambda^{-n(s+1)} \,.
\label{eq:xiandgamma}
\end{eqnarray}

\begin{figure}[!t]
\epsfxsize=3.0in
\centerline{\epsffile{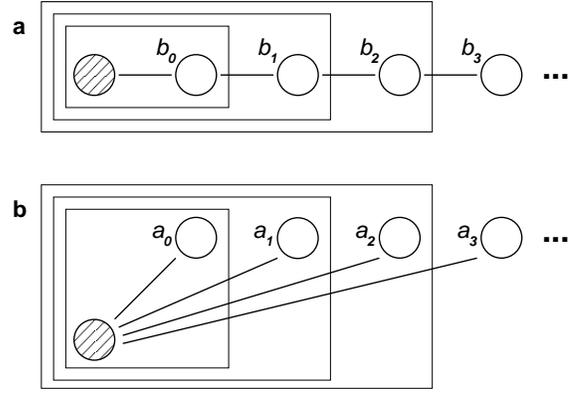}}
\caption{
(a) Structure of the spin-boson model corresponding to
eq.~(\ref{eq:hchain})
as used for the chain-NRG and (b) to eq.~(\ref{eq:hstar})
as used for the star-NRG; the boxes indicate the iterative diagonalization
scheme for both cases.
}
\label{fig:model}
\end{figure}

The structure of this Hamiltonian is sketched in  Fig.~\ref{fig:model}b:
the impurity spin couples linearly to all the bosonic degrees
of freedom $a_{n}$, in a very similar way as in the original Hamiltonian
(\ref{eq:sbm}). The bath spectral function for model
(\ref{eq:hstar}) is discrete, consisting
of $\delta$-peaks at energies $\xi_{n}$ with weight
$\propto \gamma_n^2$. Each bosonic degree of
freedom of this star-Hamiltonian is a representative of
the continuous spectrum of bosonic degrees of
freedom in the intervals $[\Lambda^{-(n+1)},\Lambda^{-n}]$.

\subsection{Chain-NRG vs. Star-NRG}

Starting from the model (\ref{eq:hstar}), there are two possible ways
to set up a numerical renormalization group procedure.
The first one (which we call ``chain''-NRG in the following)
uses the transformation of the star-Hamiltonian  (\ref{eq:hstar}) to a
semi-infinite chain:
\begin{eqnarray}
  H_{\rm c}&=& H_{\rm loc} +
   \sqrt{\frac{\eta_0}{\pi}} \frac{\sigma_{z}}{2} \left(b_{0}+b_{0}^{\dagger} \right) \nonumber\\
   &+&\sum\limits_{n=0}^{\infty} \left[ \epsilon_{n}b_{n}^{\dagger}b_{n}
         +t_{n}\left( b_{n}^{\dagger} b_{n+1}+b_{n+1}^{\dagger} b_{n}\right)
      \right]  \label{eq:hchain} \ ,
\end{eqnarray}
with $\eta_0 = \int{\rm d}x\,J(x)$.
The spin now couples to the first site of the bosonic chain
only (see Fig.~\ref{fig:model}a)
and the remaining part of the chain is characterized
by on-site energies $\epsilon_n$ and hopping parameters $t_n$,
in analogy to the fermionic NRG.
The parameters $\epsilon_n$ and $t_n$ can be calculated numerically
from a given spectral function $J(\omega)$, as
discussed in detail in Appendix \ref{app:A}.

Such a mapping from a star-Hamiltonian on a
 semi-infinite chain form is exact. It has been
used in all applications of the fermionic NRG since
the original work of Wilson \cite{Wil75}. Its generalization
to the bosonic NRG is straightforward and  has been
employed in Ref.~\onlinecite{BTV} (see also Ref.~\onlinecite{footalex}).

The structure of the Hamiltonian (\ref{eq:hchain}) is sketched in
Fig.~\ref{fig:model}a.
The boxes indicate the NRG strategy used in this case: in the first
step, a cluster containing the impurity plus the
first bath site is diagonalized. In each subsequent step, the cluster
is enlarged by one additional site and the new cluster is
diagonalized using the information obtained in the previous step.

The second possibility (which we call ``star''-NRG in the following)
is to use the Hamiltonian (\ref{eq:hstar}) {\em directly} for the iterative diagonalization.
The general idea is sketched in Fig.~\ref{fig:model}b: again, the first step of the
renormalization group procedure involves the diagonalization of a cluster
containing the impurity plus the
first bath site. The following renormalization group steps, however, are completely different
to the chain-NRG as each new bosonic site does not couple
to the previously added site but to the impurity instead.

The suggestion to use such a star-NRG for the investigation
of bosonic impurity models, such as the spin-boson model,
raises a couple of questions:

\begin{itemize}
\item[(1)] Does the star-NRG work at all?
\item[(2)] Is the star-NRG of any advantage as compared to the
    chain-NRG (apart from the simplification that we do not
    have to calculate the $\epsilon_n$ and $t_n$ of eq.~(\ref{eq:hchain}))?
\item[(3)] Why has such a star-NRG not been used in the fermionic case?
\end{itemize}

The answers to questions No.~(1) and (2) will be given further below.
Let us first discuss question No.~(3) in more detail. A fermionic star-NRG
for, say, the Kondo model would start from a Hamiltonian similar
to eq.~(\ref{eq:hstar}). The important difference in the fermionic case
is that the logarithmic discretization has to be performed for both
positive and negative frequencies. As a consequence, there
are two sets of bath operators in the star-Hamiltonian, one for
positive and one for negative frequencies:
\begin{equation}
\sum\limits_{\sigma,n=0}^{\infty}
\xi^+_{n}f_{\sigma,n+}^{\dagger}f_{\sigma,n+} +
\sum\limits_{\sigma,n=0}^{\infty}
\xi^-_{n}f_{\sigma,n-}^{\dagger}f_{\sigma,n-} \ .
\end{equation}
For a hybridization function close to particle-hole symmetry
we have $\xi^+_{n} \approx \xi^-_{n}$. This means that at each
renormalization group step one has to add {\em two} fermionic
sites (the alternative to add $f_{\sigma,n+}^{(\dagger)}$ first
and then $f_{\sigma,n-}^{(\dagger)}$, or vice versa, suffers from
violating particle-hole symmetry, if present). The Hilbert space
therefore increases by a factor of 16 in each step. It
is much more convenient to first map the star-Hamiltonian to
a chain form similar to eq.~(\ref{eq:hchain}). In this form,
only one site has to be added in each renormalization group step.

Whether such a fermionic star-NRG is of any advantage is
not clear. It might be useful for
extreme asymmetric cases, but for the cases which are usually
of interest the chain-NRG already works very well and is much easier
to implement.

Coming back to the bosonic NRG, there does not seem to be
an a priori preference for either star- or chain-NRG
because the structure of the bosonic bath is extremely asymmetric
from the outset (restricted to positive frequencies only).
To address the possible advantages of the star-NRG,
we first have to give more details of how the bosonic NRG
is implemented (for both star- and chain-NRG).

\subsection{Iterative Diagonalization and Choice of Bosonic Basis
States}
\label{sec:id}

The star-Hamiltonian $H=H_{\rm s}$ (\ref{eq:hstar})
and the chain-Hamiltonian $H=H_{\rm c}$ (\ref{eq:hchain})
can be written as a series of Hamiltonians $H_N$ ($N\ge 0$)
equal to $H$ in the limit $N\to \infty$:
\begin{equation}
    H = \lim_{N\to \infty} \Lambda^{-N} H_N \ .
\end{equation}
The $H_N$ for the star-Hamiltonian are given by
\begin{equation}
H_{N,\rm s} =
\Lambda^N \left[
H_{\rm loc} + \sum\limits_{n=0}^{N} \xi_{n}a_{n}^{\dagger}a_{n}
  + \frac{\sigma_z}{2\sqrt{\pi}}
  \sum\limits_{n=0}^{N} \gamma_{n} \left(a_{n}+a_{n}^{\dagger} \right)
\right],
\label{eq:hstarN}
\end{equation}
and for the chain-Hamiltonian by
\begin{eqnarray}
H_{N,\rm c}&=&\Lambda^N \left[ H_{\rm loc} +
   \sqrt{\frac{\eta_0}{\pi}} \frac{\sigma_{z}}{2} \left(b_{0}+b_{0}^{\dagger} \right)\right. \nonumber\\
   &+&\left.\sum\limits_{n=0}^{N} \epsilon_{n}b_{n}^{\dagger}b_{n}
         +\sum\limits_{n=0}^{N-1}
t_{n}\left( b_{n}^{\dagger} b_{n+1}+b_{n+1}^{\dagger} b_{n}\right)
      \right]  \label{eq:hchainN} \ .
\end{eqnarray}
In this notation, both $H_{0,\rm s}$ and  $H_{0,\rm c}$ correspond to
a two-site Hamiltonian with only the first site of the star or chain
coupled to the spin.

Two successive Hamiltonians are related by the following renormalization
group transformations:
\begin{eqnarray}
 & & H_{N+1,\rm s} = \Lambda H_{N,\rm s}\nonumber \\ &+&
         \Lambda^{N+1} \left[ \xi_{N+1}a_{N+1}^{\dagger}a_{N+1}
  + \frac{\sigma_z}{2\sqrt{\pi}}
      \gamma_{N+1} \! \left(a_{N+1}+a_{N+1}^{\dagger} \right)
\right] , \nonumber \\ \
\label{eq:rg-Hs}
\end{eqnarray}
and
\begin{eqnarray}
& &  H_{N+1,\rm c} = \Lambda H_{N,\rm c}\nonumber \\
       &+&
         \Lambda^{N+1} \left[
          \epsilon_{N+1}b_{N+1}^{\dagger}b_{N+1}
           + t_{N} \! \left( b_{N}^{\dagger} b_{N+1}+b_{N+1}^{\dagger} b_{N}\right)
\right] . \nonumber \\ \
\end{eqnarray}
The factor $\Lambda^{N}$ in eqs.~(\ref{eq:hstarN}) and
(\ref{eq:hchainN}) enables the direct comparison of
the low-frequency spectra of subsequent Hamiltonians and,
in particular, the discussion of fixed points as
in Sec.~\ref{sec:flow}. In contrast to the fermionic case, the factor is
$\Lambda^{N}$ instead of $\Lambda^{N/2}$ because the
energies $\xi_n$ in the star-Hamiltonian  and the $\epsilon_n$
and $t_n$ in the chain-Hamiltonian
are falling off
as $\Lambda^{-n}$, instead of the $t_n\propto\Lambda^{-n/2}$
in the fermionic case.
(This implies that a bosonic NRG calculation with discretization parameter
$\Lambda$ and a particle-hole symmetric fermionic one with $\Lambda^2$ will
have comparable energy resolution.)
Note that, in the sub-Ohmic spin-boson case,
the $\gamma_n$ are falling off {\em slower} than $\Lambda^{-n}$.
Nevertheless, the factor $\Lambda^{-n}$ is the appropriate one
for the low-energy spectra as shown in Sec.~\ref{sec:oblfp}.

The sequences of Hamiltonians (\ref{eq:hstarN}) and
(\ref{eq:hchainN}) are solved by iterative diagonalization.
In the first step, the $H_0$ are diagonalized
in a basis formed by the product states of $\sigma_z$-eigenstates
$\vert \sigma \rangle$ and a suitable basis for the first bath site
(we will describe below what we mean with `suitable basis').
We have to introduce a cutoff $N_{{\rm b}0}$ already for this
basis, but this is usually not a serious restriction as we can use fairly
large values of  $N_{{\rm b}0}\approx 500$ (in contrast to
the much lower values of $N_{\rm b}$ for the following iterations).

Given the eigenstates $\vert r \rangle_N$ of $H_{N}$
\begin{equation}
   H_{N} \vert r \rangle_N = E_N(r) \vert r \rangle_N \ \ , \ \
   r=1,\ldots N_{\rm s} \ ,
\end{equation}
with $ N_{\rm s}$ the dimension of $H_{N}$, we can construct
a basis of $H_{N+1}$:
\begin{equation}
   \vert r;s \rangle_{N+1} = \vert r \rangle_N \otimes \vert s(N+1) \rangle \ ,
\end{equation}
with $\vert s(N+1) \rangle$ a suitable basis for the added site.
In setting up the basis $\vert s(N+1) \rangle$ we are faced with two
problems not present in the fermionic case:
\begin{itemize}
  \item[(1)] The numerical approach restricts the number of basis
             states one can take into account to a
             maximum number $N_{\rm b}\approx 10-14$. The validity
             of this approximation has to be checked carefully.
  \item[(2)] A criterion for a suitable selection of $N_{\rm b}$ basis
             states out of the infinitely many states of the
             added site has to be found.
\end{itemize}

A general criterion for an ``optimal'' basis (for a given
 $N_{\rm b}$) can be formulated
as following:  find a set of $N_{\rm b}$ boson states
$\vert s(N+1) \rangle$ which give the best description of the
lowest-lying many-particle states of
$H_{N+1}$ (see also Ref.~\onlinecite{Jeckelmann}). In a variational
sense, this corresponds to finding states $\vert s(N+1) \rangle$
which give the lowest many-particle energies for a whole set
of energy levels (see also Fig.~\ref{fig:Evstheta} below).
This is certainly not a rigorous statement and we have not yet developed
a general algorithm to set up such an optimal basis. Instead we
select one of the two sets of basis states optimized for the
two stable fixed points of the spin-boson model: the
$N_{\rm b}$ eigenstates of
$b_{N+1}^\dagger b_{N+1}$ (or
$a_{N+1}^\dagger a_{N+1}$) with lowest eigenvalues as an
optimal basis for the delocalized fixed point
(Sec.~\ref{sec:obdfp}) and displaced oscillators
as optimal basis for the localized fixed point (Sec.~\ref{sec:oblfp}).

Before continuing let us point out that are no symmetries in the Hamiltonians
$H_{N,\rm s}$ and $H_{N,\rm c}$ (at least for the interesting
case of finite $\alpha$ and $\Delta$).
This is in contrast to the fermionic case\cite{Wil75,Kri80}, where we can use,
for example, the total spin and particle number as quantum numbers to
significantly reduce the size of the Hamiltonian matrices
(which would be of size $(4 N_{\rm s})^2$ in the absence of symmetries).
Consequently, in the bosonic NRG for the spin-boson model
there is only {\em one} matrix of size $(N_{\rm b} N_{\rm s})^2$ to be diagonalized
in each renormalization group step.
This results in a much simpler structure of the
NRG program, but limits the values of $N_{\rm s}$ to 100--200.

\subsection{Optimal Basis for the Delocalized Fixed Point}
\label{sec:obdfp}

Let us start from the $\alpha\!=\!0$ limit of the spin-boson
model in which two-level system and bosonic degrees of freedom are
completely decoupled;
for finite $\Delta$, the spin oscillations are undamped
and the system is in the delocalized phase from the outset.

The Hamiltonian in the original formulation (\ref{eq:sbm}) then takes
the form:
\begin{equation}
  H=-\frac{\Delta}{2}\sigma_{x} + \sum_{i} \omega_{i}
     a_{i}^{\dagger} a_{i} \ .
\end{equation}
For simplicity, the bias $\epsilon$ is set to zero.
The star-Hamiltonian in the $\alpha\!=\!0$ limit has the same structure:
\begin{equation}
  H_{\rm s}=-\frac{\Delta}{2}\sigma_{x} + \sum_{n} \xi_{n}
     a_{n}^{\dagger} a_{n} \ .
\end{equation}
From this structure, it is clear that the
$N_{\rm b}$ eigenstates of
$a_{N+1}^\dagger a_{N+1}$ with lowest eigenvalues form the
optimal basis:
\begin{equation}
\vert s(N+1) \rangle = \left\{  \vert n_{N+1} \rangle \right\}
\label{eq:basis-one} \ ,
\end{equation}
with
\begin{equation}
 a_{N+1}^\dagger a_{N+1} \vert n_{N+1} \rangle
                = n \vert n_{N+1} \rangle  \ \ , \ \
 n = 0,1,\ldots N_{\rm b}-1 \ .
\end{equation}
The reason is simply that here the many-particle energies are given
by the sum of the single-particle energies $\xi_n$.

The situation is similar in the chain-NRG, where the $\alpha\!=\!0$ limit
reads:
\begin{equation}
  H_{\rm c}=-\frac{\Delta}{2}\sigma_{x}
   + \sum\limits_{n=0}^{\infty} \left[ \epsilon_{n}b_{n}^{\dagger}b_{n}
         +t_{n}\left( b_{n}^{\dagger} b_{n+1}+b_{n+1}^{\dagger} b_{n}\right)
      \right] \label{eq:chain-dpf} \ .
\end{equation}

Here we choose for the basis $\vert s(N+1) \rangle$ the states
$\left\{  \vert n_{N+1} \rangle \right\}$ with
\begin{equation}
 b_{N+1}^\dagger b_{N+1} \vert n_{N+1} \rangle
                = n \vert n_{N+1} \rangle  \ \ , \ \
 n = 0,1,\ldots N_{\rm b}-1 \ .\label{eq:basis-onex}
\end{equation}
The difference to the basis for the star-Hamiltonian is that
the $\vert n_{N+1} \rangle$ are not eigenstates of the full
bosonic part in eq.~(\ref{eq:chain-dpf}). But in contrast
to the case of $\alpha>0$, the Hamiltonian (\ref{eq:chain-dpf})
conserves the total number of bosons; the many-particle
states with the lowest energies are then given by those
states which are constructed from the single-particle states
with the smallest boson numbers,
independent of whether a diagonal basis is chosen or not.

In our previous implementation of the bosonic NRG\cite{BTV}
we used the basis (\ref{eq:basis-onex}).
This is a suitable choice only if the many particle states of
$H_{N+1}$ with lowest energies are indeed constructed from states with small
boson number -- in other words, if the average values of
the boson numbers $\langle b_{N+1}^\dagger b_{N+1} \rangle$ are
small. This is the case when the system is close to the delocalized
and the quantum critical fixed points. However, the boson number
diverges when the system flows to the localized fixed point for
$s<1$ as discussed below.

\subsection{Optimal Basis for the Localized Fixed Point
(Displaced Oscillators)}
\label{sec:oblfp}

Here we consider the spin-boson model with zero
tunneling amplitude, $\Delta=0$. In this case, oscillations
between $\vert \uparrow \rangle$ and $\vert \downarrow \rangle$
are absent and the system is in the localized phase from the outset.

The Hamiltonian in the original formulation (\ref{eq:sbm}) then takes
the form:
\begin{equation}
H=
\sum_{i} \omega_{i}
     a_{i}^{\dagger} a_{i}
+\frac{\sigma_{z}}{2} \sum_{i}
    \lambda_{i}( a_{i} + a_{i}^{\dagger} ) \ .
\label{eq:sbmDeltazero}
\end{equation}
For simplicity, the bias $\epsilon$ is set to zero. As
the bath degrees of freedom now couple to a static spin, the
Hamiltonian can be decomposed in two sectors
$H_\uparrow$   for $\sigma_z = +1$ and
$H_\downarrow$ for $\sigma_z = -1$:
\begin{equation}
  H_\uparrow = \sum_i H_{i\uparrow} \ \ , \ \
  H_{i\uparrow} =  \omega_{i}
     a_{i}^{\dagger} a_{i} + \frac{\lambda_{i}}{2}
    ( a_{i} + a_{i}^{\dagger} ) \ ,
\end{equation}
($H_\downarrow$ accordingly). In each sector, we now have independent
bosonic degrees of freedom which can be written as:
\begin{equation}
  H_{i\uparrow} =  \omega_{i} \bar{a}_{i}^{\dagger} \bar{a}_{i} \ ,
\end{equation}
(dropping a constant term) with
\begin{equation}
    \bar{a}_{i} = a_{i} +\theta_i \ \ , \ \
   \theta_i = \frac{\lambda_{i}}{2\omega_{i}} \ .
\end{equation}
The quantities $\theta_i$ can be viewed as an effective (dimensionless)
coupling between impurity and bath mode $i$. Apparently,
this transformation corresponds to a displacement of
the oscillators $a_i$ by the value $+\theta_i$ for the $\uparrow$-sector
and $-\theta_i$ for the $\downarrow$-sector. The displacements
do not change the energies $\omega_{i}$. This means that the whole
many-particle spectrum of the bosonic bath is identical to
the one for the uncoupled bath except for the additional two-fold
degeneracy corresponding to the two sectors $\uparrow$ and $\downarrow$.

Note that for the original spin-boson model (\ref{eq:sbm}) the $\omega_i$
and $\lambda_i$ are not specified independently, only the
bath spectral function
$J(\omega)$ is given; therefore we cannot give explicit expressions
for the $\theta_i$ for eq.~(\ref{eq:sbm}).

The star-Hamiltonian eq.~(\ref{eq:hstar}) for $\Delta=0$ (and
$\epsilon=0$) takes a form similar to eq.~(\ref{eq:sbmDeltazero}).
Again we have two sectors with
 \begin{equation}
H_{\rm s\uparrow} =
\sum\limits_{n=0}^{\infty} \xi_{n}a_{n}^{\dagger}a_{n}
  + \frac{1}{2\sqrt{\pi}}
  \sum\limits_{n=0}^{\infty} \gamma_{n} \left(a_{n}+a_{n}^{\dagger} \right),
\end{equation}
($H_{\rm s\downarrow}$ accordingly). Using the same reasoning as above,
we can now write
\begin{equation}
H_{\rm s\uparrow} =
\sum\limits_{n=0}^{\infty} \xi_{n}\bar{a}_{n}^{\dagger}\bar{a}_{n} \ , 
\end{equation}
with
\begin{equation}
   \bar{a}_{n} =  a_{n} +\theta_n \ \ , \ \
   \theta_n = \frac{\gamma_{n}}{2\sqrt{\pi}\xi_{n}} \ .
\label{eq:theta-star}
\end{equation}
The values of $\gamma_{n}$ and $\xi_{n}$ are given in
eq.~(\ref{eq:xiandgamma}) so we obtain
\begin{equation}
 \theta_n \propto \Lambda^{n(1-s)/2} \propto \xi_n^{(s-1)/2} \ .
\label{eq:shift-star}
\end{equation}
Written in terms of energy $\omega$ we therefore have:
\begin{equation}
  \theta(\omega) \propto \omega^{(s-1)/2} \ .
\end{equation}
This result is rather interesting: for sub-Ohmic baths, $s<1$, the
shift $\theta_n$ grows exponentially with $n$.
However, in the super-Ohmic case the shift goes to zero in the
low-energy limit ($n\to\infty$), and it is energy-independent for the Ohmic case.
Technically, the coupling to the impurity can be viewed as a
relevant (irrelevant) perturbation of the discretized spin-boson
model for $s<1$ ($s>1$) and as a marginal perturbation in the Ohmic
case.
Thus, in the Ohmic and super-Ohmic case the effective coupling
$\theta(\omega)$ does {\em not} diverge as $\omega\to 0$
even in the extreme localized case of $\Delta=0$.
Therefore, numerical problems associated with a diverging effective
coupling are only expected in the sub-Ohmic case.

Coming back to the iterative {\em numerical} diagonalization
of the star-Hamiltonian, it is now clear that a simple basis
as in (\ref{eq:basis-one}) can be far from the optimal choice.
If we stay in the original basis constructed from the
lowest eigenstates of $a_n^\dagger a_n$,
we need more and more basis states to describe the lowest
eigenstates of the displaced oscillators.

On the other hand, it is clear how to construct the
optimal basis for $\vert s(N+1) \rangle$ at least
for the $\Delta\!=\!0$ case.
For the sectors $\uparrow\!/\!\downarrow$ we simply take oscillator
states with displacements $+\theta_{N+1}/-\theta_{N+1}$. As we need
a single basis for both sectors, these states have to be orthogonalized
first; this will be discussed in more detail in
Appendix \ref{app:B}.

The displaced oscillator states can also be used to diagonalize
the chain-Hamiltonian (\ref{eq:hchain}) for $\Delta=0$.
For a given iteration number $N$, the $H_{N,\rm c}$ for the
$\uparrow$-sector reads:
\begin{eqnarray}
H_{N,\rm c \uparrow}&=&\Lambda^N \left[
   \frac{1}{2} \sqrt{  \frac{\eta_0}{\pi}} \left(b_{0}+b_{0}^{\dagger} \right)\right. \nonumber\\
   &+&\left.\sum\limits_{n=0}^{N} \epsilon_{n}b_{n}^{\dagger}b_{n}
         +\sum\limits_{n=0}^{N-1}
t_{n}\left( b_{n}^{\dagger} b_{n+1}+b_{n+1}^{\dagger} b_{n}\right)
      \right] \ .
\end{eqnarray}
Introducing displaced oscillators
\begin{equation}
  \bar{b}_n = b_n + \theta_n(N) \ ,
\end{equation}
we again have a diagonal form
\begin{equation}
H_{N,\rm c \uparrow} = \Lambda^N
\sum\limits_{n=0}^{N} \epsilon_{n}\bar{b}_{n}^{\dagger}\bar{b}_{n} \ .
\end{equation}
The displacements $ \theta_n(N)$ can be calculated numerically for
any given set of $\{\epsilon_{n}\}$ and $\{t_{n}\}$.
For fixed $N$ they show the same qualitative behavior as
the $\theta_n$ for the star-Hamiltonian:
\begin{equation}
   \vert \theta_n(N)\vert \propto \Lambda^{n(1-s)/2} \ .
\end{equation}
It turns out,
however, that the $\theta_n(N)$ depend on {\em both} $n$ and $N$
with significant deviations from the exponential form for
$n$ close to $N$.
This has important
consequences for the use of displaced oscillators as basis states
in the chain-NRG. Let us assume that we used $\pm\theta_N(N)$
to construct the basis for $H_{N,\rm c}$. Adding the site $N+1$
introduces a significant change in the displacement
$ \theta_N(N) \to \theta_N(N+1)$. One possible solution to
this problem is to anticipate the coupling to the site $N+1$
by adding a static displacement term, which is subtracted again
in the next step. Such an approach gives correct results for
the chain-NRG when we set $\Delta=0$. We did not, however, succeed
in implementing the displaced oscillator idea for the general
case of {\em finite} $\Delta$ in the chain-NRG.
So far, this strategy only works for the star-NRG as described in
the following subsection.

\subsection{General Strategy of the Bosonic NRG}
\label{sec:genstrat}

In the preceding subsections we have described various options
of how to set up the bosonic NRG. We have introduced both a
star- and a chain-representation of the spin-boson model
and we discussed two possibilities for choosing a basis for
the added site: eigenstates of $b_{N+1}^\dagger b_{N+1}$
(or $a_{N+1}^\dagger a_{N+1}$) as in Sec.~\ref{sec:obdfp}
   or displaced oscillators
as in Sec.~\ref{sec:oblfp}.

Now we want to discuss how we actually proceed with the bosonic NRG:
how do we decide between the different options described above?

As a starting point we choose the chain-NRG using
eigenstates of $b_{N+1}^\dagger b_{N+1}$
(the basis denoted by $\vert n_{N+1} \rangle$)
as the simplest possible basis.
This approach has been used for all the results shown in
Ref.~\onlinecite{BTV}.
From the discussion in Secs.~\ref{sec:obdfp} and \ref{sec:oblfp}
we anticipate that this choice of the basis is reasonable
for
\begin{itemize}
  \item all parameters in the super-Ohmic case,
  \item the Ohmic case provided the coupling $\alpha$ is not too large,
  \item and the sub-Ohmic case provided the system is close to
        the delocalized fixed point.
\end{itemize}

On the other hand, it is clear that there will be problems
when the system flows to the localized fixed point
in the sub-Ohmic case.
The situation in the crossover regions and close to the quantum critical points
needs to be checked numerically: it turns out that the critical fixed points
for all $0<s<1$ can be reached using the $\vert n_{N+1} \rangle$ basis.

There is a simple criterion to decide when the basis
$\vert n_{N+1} \rangle$ is sufficient. Consider the
expectation value $n_{N+1}(N_{\rm b}) = \langle
b_{N+1}^\dagger b_{N+1} \rangle$ for the cluster
after adding the site $N+1$, calculated for a temperature
of the order of the level spacing at this NRG step.
This quantity can be obtained numerically
up to values of $N_{\rm b}\approx 14$.  If the lowest eigenvalues of
$b_{N+1}^\dagger b_{N+1}$ are a good choice for describing the
lowest eigenstates of $H_{N+1}$, then  $n_{N+1}(N_{\rm b})$ should
be small and rapidly saturate with increasing $N_{\rm b}$. If,
on the other hand, we identify that $n_{N+1}(N_{\rm b})$ does not
saturate but increases with   $N_{\rm b}$, then we certainly have
to abandon the basis $\vert n_{N+1} \rangle$ and use a different
`optimized' basis.

\begin{figure}[!t]
\epsfxsize=2.8in
\centerline{\epsffile{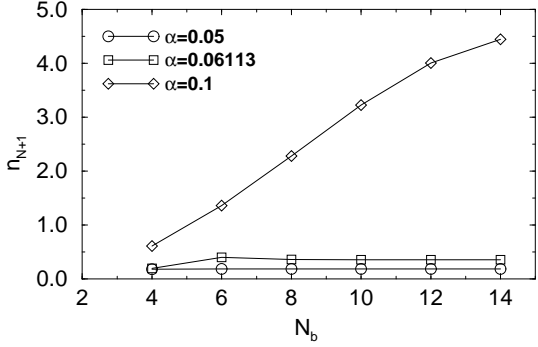}}
\caption{
Dependence of the expectation value
$n_{N+1} = \langle
b_{N+1}^\dagger b_{N+1} \rangle$ on the number of basis
states $N_{\rm b}$ for a chain-NRG calculation. The values
of $n_{N+1}(N_{\rm b})$ quickly saturate for
$\alpha = 0.05<\alpha_{\rm c} $ and $\alpha_{\rm c} = 0.06113$
whereas no saturation is observed for $\alpha = 0.1 > \alpha_{\rm c}$.
(Parameters are $s=0.6$, $N=20$, $\Lambda=2.0$, $N_{\rm s}=60$, $\Delta=0.01$).
}
\label{fig:nb}
\end{figure}

This behavior is shown in Fig.~\ref{fig:nb} where we show results from the
chain-NRG for a sub-Ohmic bath ($s=0.6$), $\Delta=0.01$ and three
values of $\alpha$ in the vicinity of the quantum phase transition. For
$\alpha<\alpha_{\rm c}$ and  $\alpha=\alpha_{\rm c}$ we indeed
find a rapid saturation of $n_{N+1}(N_{\rm b})$ whereas
no saturation  (at least up to $N_{\rm b}=14$)
is observed for $\alpha>\alpha_{\rm c}$.

The behavior of  $n_{N+1}(N_{\rm b})$ for  $\alpha>\alpha_{\rm c}$
can be easily understood from the discussion of Sec.~\ref{sec:oblfp}:
as the system is flowing to the localized fixed point corresponding
to the effective $\Delta$ approaching zero, we have to use
properly displaced oscillators as a basis. The increase of
$n_{N+1}(N_{\rm b})$ just means that we need more and more states in
the undisplaced basis to describe the lowest eigenstates of
$H_{N+1}$.

In this case, the use of displaced oscillators as introduced
in Sec.~\ref{sec:oblfp} is much more appropriate. Note, however, that
the shifts $\theta_n$ eq.~(\ref{eq:theta-star}) can only be defined
from the outset for the $\Delta\!=\!0$ case. For any finite
$\Delta$, the system evolves according to the iterative
diagonalization. If the system turns out to flow to the
localized fixed point, we have to use {\em effective}
displacements $\theta_n$ to set up the basis. These displacements
have to be extracted numerically from the renormalization group
calculation and are different (for finite $\Delta$) from the
$\theta_n$ given in eq.~(\ref{eq:theta-star}).

\begin{figure}[!t]
\epsfxsize=2.9in
\centerline{\epsffile{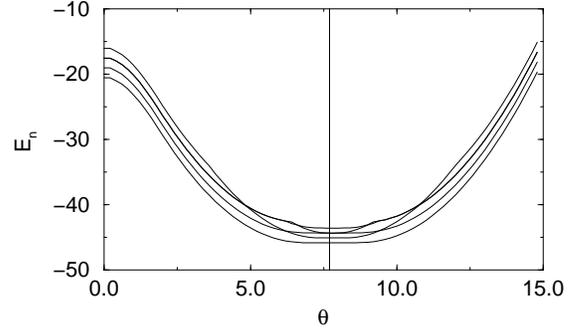}}
\caption{
Dependence of the energies of the lowest eigenstates
of $H_{N+1}$ on the displacement $\theta$ used for constructing
the basis for the new degree of freedom added in each iteration
step. All levels shown here have their minimum at the same
value $\theta = \theta^\ast$ (indicated by the vertical line)
which is the optimal value
for setting up the basis. The parameters for this calculation
are: $s=0.2$, $\Delta=1.0$, $\alpha=0.25 > \alpha_{\rm c}$.
}
\label{fig:Evstheta}
\end{figure}

Figure \ref{fig:Evstheta} describes the general strategy to determine
the optimal values of the displacements. The low-energy spectrum of
$H_{N+1}$ is calculated for a whole set of $\theta$-values. According to
the discussion in Sec.~\ref{sec:id}, we identify the optimal $\theta$ as the
one which gives the lowest eigenenergies in $H_{N+1}$. This value
is indicated by the vertical line in Fig.~\ref{fig:Evstheta}, which
shows results for the sub-Ohmic case and parameters close to the
localized fixed point. There
is a plateau in the energy levels close to the optimal value which
means that a slight variation of the $\theta$ affects the lowest
energies only very weakly. Note that $E_n(\theta)=E_n(-\theta)$,
therefore a maximum at $\theta=0$. The corresponding figure for
parameters close to the delocalized fixed point (not shown here)
gives a minimum of the many-particle levels at $\theta=0$.
For further details
of this procedure, see Appendix \ref{app:B}.

The data of Fig.~\ref{fig:Evstheta} are calculated using the star-NRG
formulation. Although a similar figure can be generated using the chain-NRG,
we are facing the (so far unsolved) problem discussed in \ref{sec:oblfp}:
adding a new site changes the optimal displacements for the previous
iterations.
For this reason, all the results in this paper using a basis of displaced
oscillators are calculated within a star-NRG representation.

\subsection{Diagonalization and Truncation}
\label{sec:dt}

To conclude Sec.~\ref{sec:bosonicNRG}, let us briefly discuss the
remaining technical steps necessary to complete the iterative
diagonalization. For a given basis, we first set up the Hamiltonian
matrices
\begin{equation}
   H_{N+1}(rs,r^\prime s^\prime) =\,
   _{N+1}\langle r;s \vert H_{N+1} \vert r^\prime; s^\prime
      \rangle_{N+1} \ .
\end{equation}
For both chain- and star-formulation of the NRG, the matrices
can be written as a sum of three parts:
\begin{equation}
   H_{N+1}(rs,r^\prime s^\prime) = H_{N+1}^{(1)} +
H_{N+1}^{(2)} + H_{N+1}^{(3)} \ ,
\label{eq:threeparts}
\end{equation}
with
\begin{eqnarray}
  H_{N+1}^{(1)}(rs,r^\prime s^\prime)
   & = & \Lambda\  _{N+1}\langle r;s \vert H_{N} \vert r^\prime; s^\prime
      \rangle_{N+1} \nonumber \\
    & = & \Lambda E_N(r) \delta_{rr^\prime} \delta_{ss^\prime} \ ,
\end{eqnarray}
for both chain- and star-formulation and
\begin{eqnarray}
&&  H_{N+1,{\rm s}}^{(2)}(rs,r^\prime s^\prime) \nonumber \\
&&    =  \Lambda^{N+1} \xi_{N+1}\  _{N+1}\langle r;s \vert
         a_{N+1}^{\dagger}a_{N+1}  \vert r^\prime; s^\prime
      \rangle_{N+1} \nonumber \\
&&    =   \Lambda^{N+1} \xi_{N+1}  \delta_{rr^\prime} \
        \langle s(N+1) \vert  a_{N+1}^{\dagger}a_{N+1} \vert
        s^\prime(N+1) \rangle \ ,
\end{eqnarray}
(with operators $a$ replaced by $b$ for  $H_{N+1,{\rm c}}^{(2)}$).

The third term takes the following form for the star-NRG
\begin{eqnarray}
&&  H_{N+1,{\rm s}}^{(3)}(rs,r^\prime s^\prime) \nonumber \\
&& =  \Lambda^{N+1}  \frac{ \gamma_{N+1}}{2\sqrt{\pi}}
         \  _{N+1}\langle r;s \vert
         \sigma_z  (a_{N+1}+a_{N+1}^{\dagger} )
           \vert r^\prime; s^\prime
      \rangle_{N+1} \nonumber \\
&& =  \Lambda^{N+1}  \frac{ \gamma_{N+1}}{2\sqrt{\pi}}
         \  _{N}\langle r \vert
         \sigma_z \vert r^\prime  \rangle_{N} \nonumber \\
&&\ \ \times
         \langle s(N+1) \vert a_{N+1} +  a_{N+1}^{\dagger} \vert
        s^\prime(N+1) \rangle \ ,
\label{eq:H3s}
\end{eqnarray}
and for the chain-NRG
\begin{eqnarray}
&&  H_{N+1,{\rm c}}^{(3)}(rs,r^\prime s^\prime) \nonumber \\
&& =  \Lambda^{N+1}  t_{N}  \  _{N+1}\langle r;s \vert
           b_{N}^{\dagger} b_{N+1}+ h.c.  \vert r^\prime; s^\prime
      \rangle_{N+1} \nonumber \\
&& =   \Lambda^{N+1}  t_{N}  \,  _{N}\langle r \vert
           b_{N}^{\dagger} \vert r^\prime  \rangle_{N}
          \langle s(N+1) \vert  b_{N+1} \vert
        s^\prime(N+1) \rangle  \nonumber \\
&&\ \ \  + h.c. \ .
\label{eq:H3c}
\end{eqnarray}
All matrix elements of the form
$ \langle s(N+1) \vert \ldots \vert   s^\prime(N+1) \rangle$
can be further simplified once the basis
$\vert   s(N+1) \rangle$ is given. Similar to the fermionic case,
the matrix element
$_{N}\langle r \vert
           b_{N}^{\dagger} \vert r^\prime  \rangle_{N}$ appearing in
the chain-NRG eq.~(\ref{eq:H3c}) can be written in terms of the
unitary matrices necessary to diagonalize $H_N$. The matrix elements
$ _{N}\langle r \vert
         \sigma_z \vert r^\prime  \rangle_{N}$ in
eq.~(\ref{eq:H3s}), however, have to be calculated iteratively.
(The technical details are very similar to the fermionic case,
see Refs.~\onlinecite{Wil75,Kri80}).

With $N_{\rm s}$ the dimension of $H_N$ and $N_{\rm b}$ the number of
basis states in $\vert   s(N+1) \rangle$, we then arrive at
a {\em single} $(N_{\rm s}\cdot N_{\rm b})\times(N_{\rm s}\cdot N_{\rm b})$
matrix for $H_{N+1}(rs,r^\prime s^\prime)$. This matrix
can be diagonalized using standard routines. From this we obtain
the unitary matrices  $U_{N+1}(rs,\bar{r})$ and the spectrum
of eigenenergies $E_{N+1}(\bar{r})$ so that
\begin{equation}
   H_{N+1} \vert \bar{r} \rangle_{N+1} =
E_{N+1}(\bar{r})  \vert \bar{r} \rangle_{N+1} \ , \
\bar{r} = 1,\ldots N_{\rm s}\cdot N_{\rm b} \ .
\end{equation}
In contrast to the fermionic case, no symmetries can be taken into
account to separate the matrix  $H_{N+1}(rs,r^\prime s^\prime)$ into
smaller sub-matrices.

The dimension of $H_{N+1}$ now has to be reduced from
 $N_{\rm s}\cdot N_{\rm b}$ to  $N_{\rm s}$ to allow for a numerical
calculation with computation time growing only linearly with $N$.
This is achieved with the usual truncation scheme where only the
lowest  $N_{\rm s}$ eigenstates of  $H_{N+1}$ are kept (for
the fermionic case see Refs.~\onlinecite{Wil75,Kri80}). These states form the
basis  $\vert r \rangle_{N+1}$ for the next step and the iteration
continues.

The calculation of correlation functions, such as the spin-spin
correlation function $C(\omega)$ in Sec.~\ref{sec:dyn},
requires the calculation
of additional matrix elements $ _{N}\langle r \vert
         \hat{A} \vert r^\prime  \rangle_{N}$.
For more details see Sec.~\ref{sec:dyn}.

\section{Flow and Fixed Points}
\label{sec:flow}
The iterative numerical diagonalization of the spin-boson model
as described in the previous section gives a sequence of
many-particle levels $E_N(r)$ ($r=1,\ldots N_{\rm s}$). Due
to the logarithmic discretization, these energies fall off
as $E_N(r) \propto \Lambda^{-N}$. NRG flow diagrams can then
be constructed by plotting $\Lambda^{N} E_N(r)$ versus iteration
number $N$.

In this section we focus on those issues which can
be directly inferred from the NRG flow diagrams: the appearance
of fixed points, the crossover between different fixed
points at finite energy or temperature, and quantum phase
transitions between the fixed points.
Subsections \ref{sec:3A}-\ref{sec:3C} deal with the
Ohmic spin-boson model; here we also address the issue
of convergence. In subsection \ref{sec:3D} we
investigate those features connected to the flow of energy levels
which are specific for the sub-Ohmic case.

All results are calculated for cutoff energy
$\omega_c=1$ and bias $\epsilon=0$;
we employ NRG parameter values of $\Lambda=1.8-3.2$, $N_{\rm b0} = 100$,
$N_{\rm b}=4-14$, $N_{\rm s}=30-120$.

\subsection{Fixed Points}
\label{sec:3A}

Let us first concentrate on results from the
chain-NRG for the Ohmic case, $s=1$, and various values of
$\Delta$ and $\alpha$.

\begin{figure}[!t]
\epsfxsize=3.0in
\centerline{\epsffile{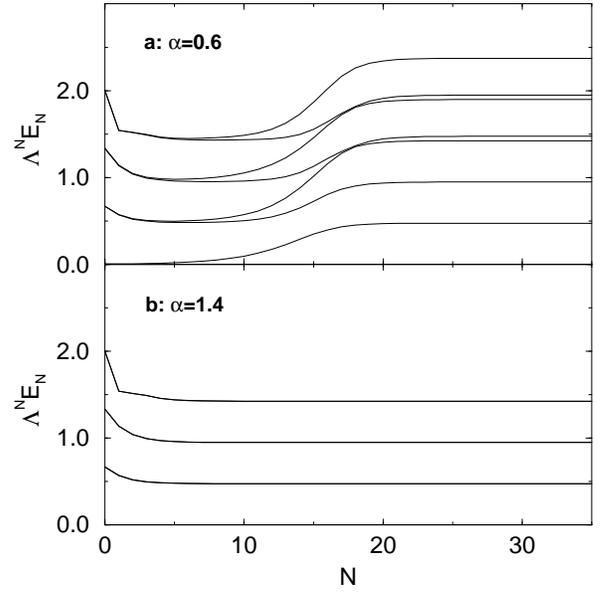}}
\caption{
Flow diagrams calculated with the chain-NRG for the parameters
$s=1$ (Ohmic case), $\omega_c=1$, $\epsilon=0$, $\Delta=0.01$,
and  $\alpha=0.6$ in (a) and  $\alpha=1.4$ in (b). The
NRG parameters are $N_{\rm s}=100$,
$N_{\rm b}=8$, and $\Lambda=2.0$.
}
\label{fig:3_1}
\end{figure}

Fig.~\ref{fig:3_1} shows two NRG flow diagrams for $\Delta=0.01$
and two values for the coupling: $\alpha=0.6$ in Fig.~\ref{fig:3_1}a
and $\alpha=1.4$ in Fig.~\ref{fig:3_1}b. In these diagrams, the rescaled
many-particle energies $\Lambda^{N} E_N(r)$ are plotted
versus the iteration number $N$ with the ground state energy subtracted.
Another difference to the fermionic case (apart from the
different prefactor $\Lambda^{N}$ instead of $\Lambda^{N/2}$)
is the absence of an even-odd effect: in the fermionic case,
the many-particle spectrum usually oscillates between two sets
of energy levels (so that it is more appropriate to speak of a
limit cycle than of a fixed point). Plotting the many-particle
spectrum either for even or for odd iteration numbers only then gives
the flow diagrams as shown in, for example, Refs.~\onlinecite{Wil75,Kri80}.

In our bosonic NRG calculations, we can follow the flow typically
up to $N=60$ (corresponding to $T\approx 10^{-20}$
for $\Lambda=2.0$),
then we observe an unphysical runaway which is
due to the accumulation of numerical errors in the course of the
iteration.
As the runaway scale depends on the numerical precision used in the
program code, it can be shifted to lower temperatures if needed.

The flow diagrams of Fig.~\ref{fig:3_1}
show the existence of two different fixed
points: the delocalized fixed point for small $\alpha$
(see Fig.~\ref{fig:3_1}a,
$N>20$) and the localized fixed point for  large  $\alpha$ 
(see Fig.~\ref{fig:3_1}b,
$N>6$). These two fixed points are stable and the quantum phase transition
between them is discussed further below.

If the value of $\Delta$ is small enough (as in Fig.~\ref{fig:3_1}a)
the system is close to the localized
fixed point in an intermediate range
($4<N<8$ in Fig.~\ref{fig:3_1}a) even for $\alpha$ values
below the critical
coupling  $\alpha_{\rm c}$.
This has direct consequences for thermodynamic properties in the
corresponding temperature range (see, for example,
Fig.~\ref{fig:4_1}). However, the vicinity to the localized
fixed point does not imply localization in the sense that
a system initially prepared with the impurity spin in
one specified direction remains in this spin state under
time evolution. For any finite temperature, thermal
excitations destroy localization (see Ref.~\onlinecite{Leggett}).

In Fig.~\ref{fig:3_1}a, we also observe a crossover from the
localized to the delocalized fixed point which
takes place at $N\approx 10-20$.
The corresponding crossover scale, $T^\ast$
(which -- in the Ohmic case -- is equivalent to the renormalized tunnel
splitting $\Delta_{\rm r}$ up to a prefactor),
will be discussed in subsection \ref{sec:3C}.

The flow diagram of Fig.~\ref{fig:3_1}a is similar to the one obtained
in Ref.~\onlinecite{Costi98}
(Fig.~1 in Ref.~\onlinecite{Costi98}), where the mapping of the spin-boson
model to the anisotropic Kondo model was employed. The structure of
the many-particle levels, however, cannot be directly compared as they
reflect the type of bath used in the NRG approach (bosonic in our case,
fermionic in Ref.~\onlinecite{Costi98}).

The spectrum of the delocalized
fixed point (Fig.~\ref{fig:3_1}a for $N>20$)
is identical to the spectrum of a spin-boson model with zero coupling
between spin and bosons ($\alpha=0$). The $H_N$ for the chain-NRG then
take the form:
\begin{eqnarray}
H_{N,\rm c}&=&\Lambda^N \left[ H_{\rm loc}+
             \sum\limits_{n=0}^{N} \epsilon_{n}b_{n}^{\dagger}b_{n} \right.
   \nonumber\\
   &+&\left.
         \sum\limits_{n=0}^{N-1}
t_{n}\left( b_{n}^{\dagger} b_{n+1}+b_{n+1}^{\dagger} b_{n}\right)
      \right] 
\end{eqnarray}
In this Hamiltonian, impurity and bath degrees of freedom
 are completely decoupled and can be diagonalized separately.
The spectrum of the impurity part
($H_{\rm loc}= -\Delta \sigma_x/2$) is non-degenerate.
The bath part is that of a free chain of bosons
with $N+1$ sites which can be diagonalized exactly:
\begin{eqnarray}
\sum\limits_{n=0}^{N} \epsilon_{n}b_{n}^{\dagger}b_{n}
         +\sum\limits_{n=0}^{N-1}
t_{n}\left( b_{n}^{\dagger} b_{n+1}+b_{n+1}^{\dagger} b_{n}\right)
= \sum \limits_{n=0}^{N} \bar{\omega}_n \bar{b}_{n}^{\dagger}\bar{b}_{n}
\ . \nonumber\\
\label{eq:baromega}
\end{eqnarray}

\begin{figure}[!t]
\epsfxsize=3.0in
\centerline{\epsffile{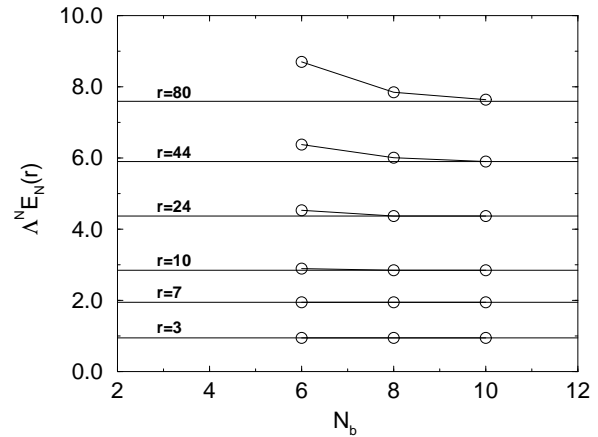}}
\caption{
Comparison between the fixed-point spectra
for the delocalized fixed point
calculated with the chain-NRG (circles) for various values
of  $N_{\rm b}$ and the
fixed point spectra constructed from the single particle levels
$\bar{\omega}_n$ in eq.~(\ref{eq:baromega}) (solid lines) for
a selection of states $E_N(r)$.
The parameters are
$s=1$, $\Delta=1.0$,
and  $\alpha<\alpha_{\rm c}$. The
NRG parameters are $N_{\rm s}=100$,  and $\Lambda=2.0$.
}
\label{fig:3_2}
\end{figure}

Figure \ref{fig:3_2} shows a comparison between the fixed point spectra
for the delocalized fixed point
calculated with the chain-NRG (circles) and the
fixed-point spectra constructed from the single particle levels
$\bar{\omega}_n$ in eq.~(\ref{eq:baromega}) (solid lines).
The NRG data are calculated for different $N_{\rm b}$. The
agreement is very good for the first few excitations
already for $N_{\rm b}\approx 6$, while a larger value
of $N_{\rm b}$ is required to correctly reproduce the excitations
at higher energies.

While the delocalized fixed point is reached for $\alpha$ smaller than a
critical $\alpha_{\rm c}(\Delta)$, the system is in the localized
phase for all $\alpha>\alpha_{\rm c}(\Delta)$.
The localized phase is characterized by a (renormalized) tunneling amplitude
$\Delta_{\rm r}=0$ and a two-fold degenerate ground state.
In the language of the (perturbative) renormalization group\cite{Leggett,AYH}
the localized phase corresponds to a {\em line} of fixed points,
parametrized by $\alpha$.
Interestingly, the fixed-point value of $\alpha$ does {\em not}
influence the eigenenergies of the many-body fixed-point Hamiltonian,
but only its eigenstates, see the discussion in
Sec.~\ref{sec:bosonicNRG}.
Thus the NRG level spectrum in the entire localized phase is {\em identical}
to the one for the delocalized fixed point, apart from an additional
two-fold degeneracy of all many-particle levels.
This feature can be clearly seen in Fig.~\ref{fig:3_1}.
[Of course, the approach to the localized fixed point depends
on the particular value of $\alpha$,
consequently the NRG flow on intermediate scales will be different
for different $\alpha>\alpha_{\rm c}(\Delta)$.]

\subsection{Critical Coupling and Convergence}
\label{sec:3B}

The results shown in Fig.~\ref{fig:3_1} indicate the well-known transition
between the localized and delocalized fixed points at a critical
$\alpha_{\rm c}(\Delta)$ \cite{Leggett,Weiss}. Due to the Kosterlitz-Thouless
nature of this
transition, the fixed point at $\alpha=\alpha_{\rm c}(\Delta)$ is
{\em not} a new fixed point, but belongs to the localized phase
instead.

On approaching the transition from the delocalized side, we find, as
expected, that the crossover scale vanishes as\cite{expnote}
$\ln T^\ast\propto 1/(\alpha_{\rm c} - \alpha)$, see
Fig.~\ref{fig:3_8} below.
We use this dependence to determine the value of $\alpha_{\rm c}$
from numerically calculated data for $T^\ast(\alpha)$ via a non-linear
fit.
[Note that on the localized side of the transition a low-energy scale
only shows up in the flow towards the fixed point, i.e., in corrections
to the fixed-point values of observables;
thus the critical $\alpha_{\rm c}(\Delta)$ is easier obtained via extrapolation
from the {\em delocalized} side.]

\begin{figure}[!t]
\epsfxsize=2.8in
\centerline{\epsffile{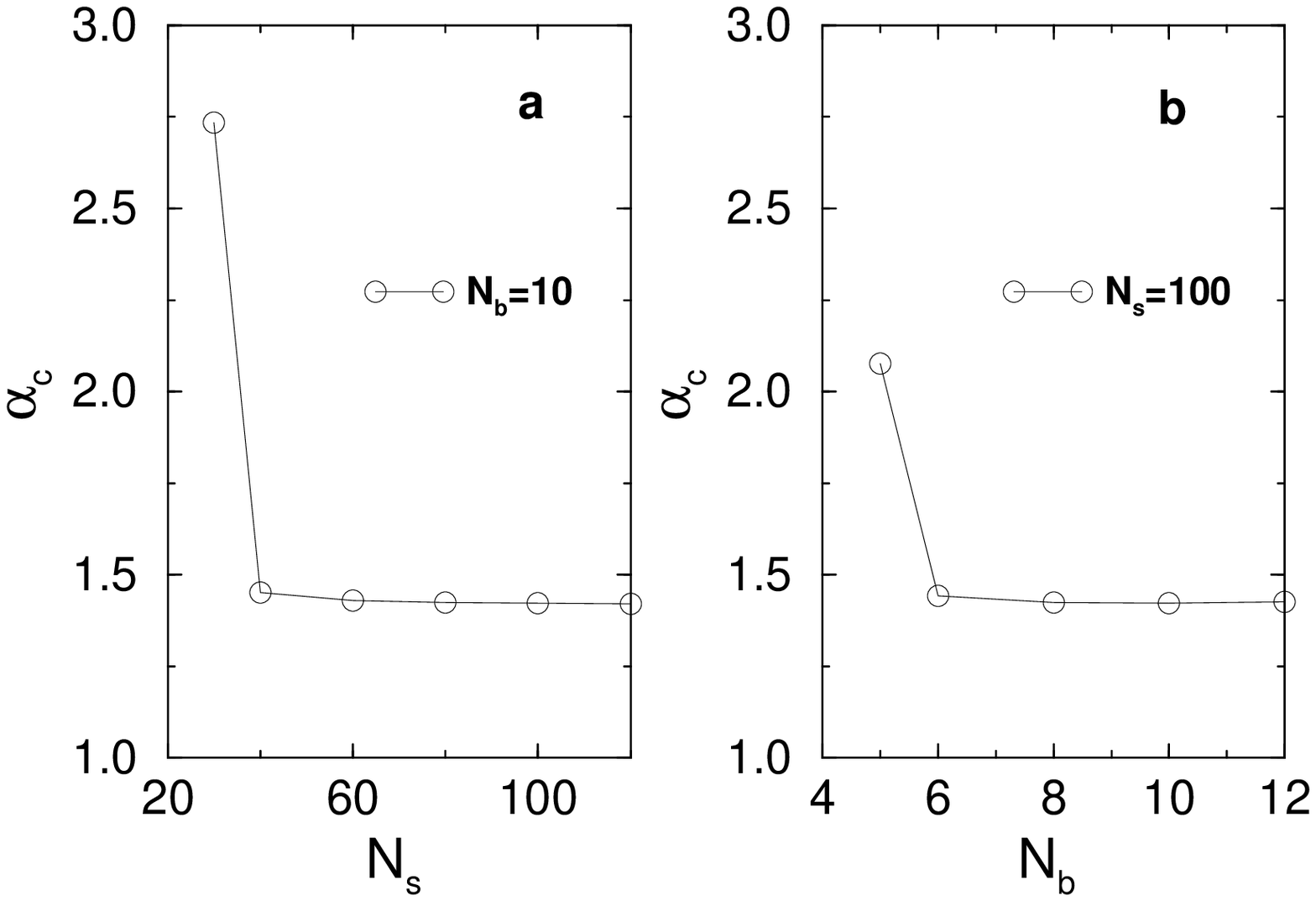}}
\epsfxsize=2.8in
\centerline{\epsffile{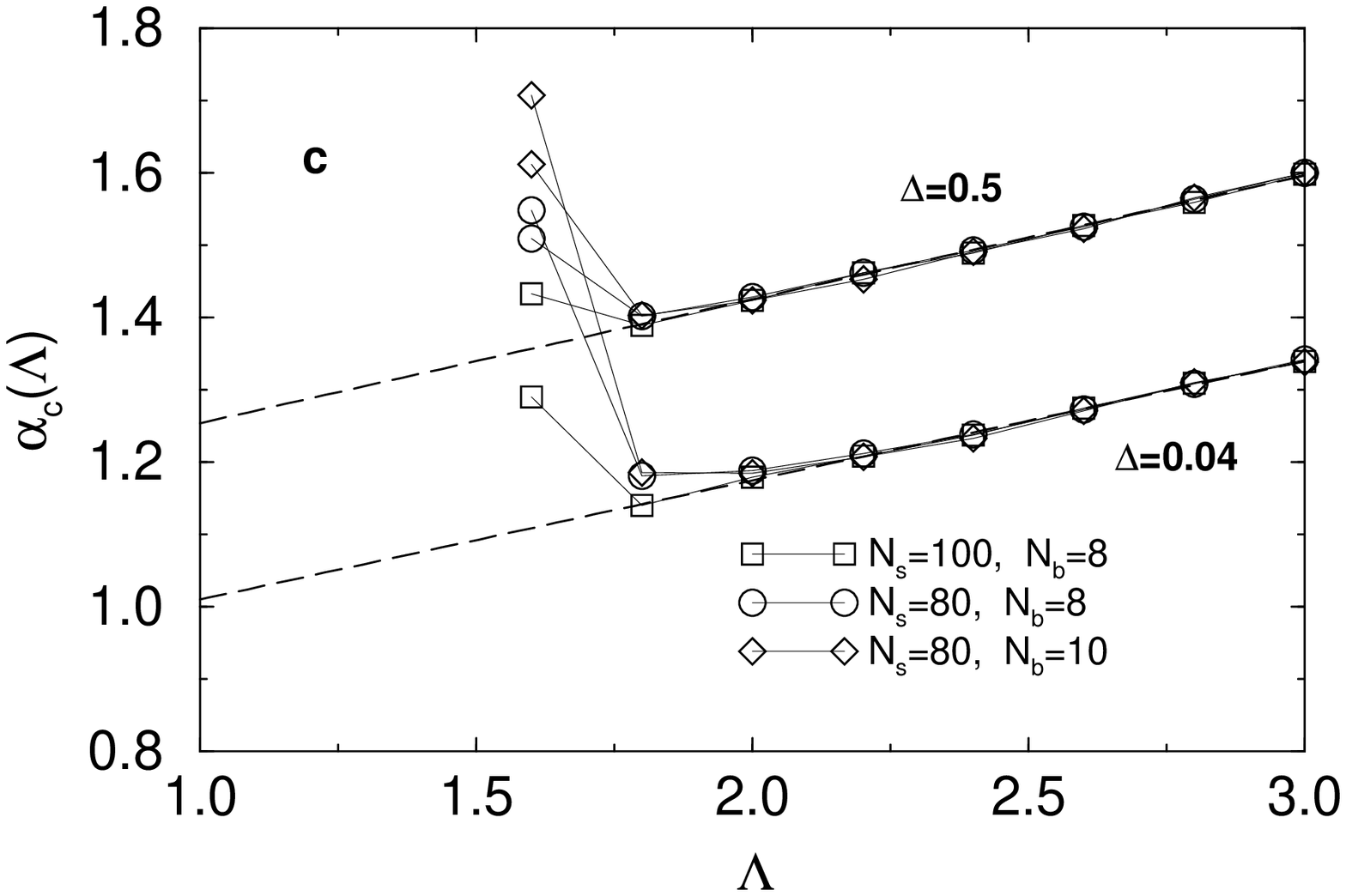}}
\caption{
Dependence of the critical coupling  $\alpha_{\rm c}$
on the  NRG parameters $N_{\rm s}$, $N_{\rm b}$, and $\Lambda$
for the Ohmic case; (a) dependence on $N_{\rm s}$ for fixed
$N_{\rm b}=10$; (b) dependence on $N_{\rm b}$ for fixed
$N_{\rm s}=100$ ($\Lambda=2.0$ for both (a) and (b));
(c)
$\Lambda$-dependence of $\alpha_{\rm c}$ for two values of $\Delta$,
and various NRG parameters
$N_{\rm b}$ and $N_{\rm s}$. The dashed lines are  linear fits to
the $N_{\rm b}=8$ and $N_{\rm s}=100$ data in the range
$1.8\le \Lambda \le 3$.
}
\label{fig:3_4}
\end{figure}

As already discussed in Ref.~\onlinecite{BTV} (see Fig.~2 in Ref.~\onlinecite{BTV}),
the  value of $\alpha_{\rm c}$ also depends on the NRG parameters
$\Lambda$, $N_{\rm b}$, and $N_{\rm s}$. Figures \ref{fig:3_4}a-c show the
characteristic dependence for the Ohmic case, $s=1$, and
two values of $\Delta$ in Fig.~\ref{fig:3_4}c.
Keeping $\Lambda$ fixed, we observe a rapid convergence of $\alpha_{\rm c}$
with increasing $N_{\rm s}$ (Fig.~\ref{fig:3_4}a) and
$N_{\rm b}$ (Fig.~\ref{fig:3_4}b).
Note that we did not observe a transition to the delocalized phase
for $N_{\rm b}\le 4$, even for very large values of $\alpha$.
As expected from the iterative diagonalization scheme,
the values of $N_{\rm s}$ and $N_{\rm b}$ necessary for convergence
increase with decreasing $\Lambda$ (see Fig.~\ref{fig:3_4}c).
The converged data for $\alpha_{\rm c}(\Lambda)$
show a linear $\Lambda$ dependence in the range
$1.8<\Lambda<3$, with a deviation of about 15\% at
$\Lambda = 2$ from the extrapolated $\Lambda\to 1$ value.

We find that the slope in $\alpha_{\rm c}(\Lambda)$ is independent of $\Delta$
which is connected to fact that the logarithmic discretization
systematically underestimates the spectral weight contained
in $J(\omega)$ (for a discussion of this point in the fermionic
case, see eq.~(5.42) in Ref.~\onlinecite{Kri80}; for the soft-gap
Anderson model see Fig.~4 in Ref.~\onlinecite{BGLP}).

\begin{figure}[!t]
\epsfxsize=2.8in
\centerline{\epsffile{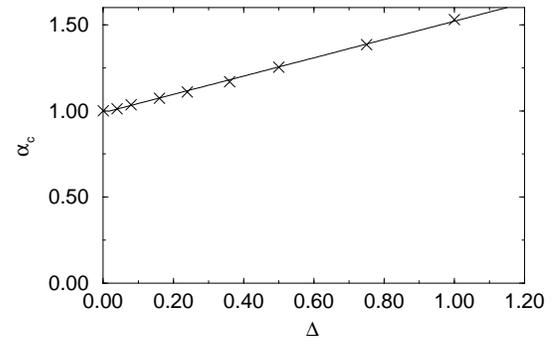}}
\caption{
Dependence of the extrapolated value $\alpha_{\rm c}(\Lambda\to 1)$
on the parameter $\Delta$ for the Ohmic case $s=1$. The
crosses are the numerical data and the solid line is a linear
fit which gives $\alpha_{\rm c}(\Delta)=0.99 + 0.53\Delta$.
The NRG parameters are $N_{\rm s}=100$ and
$N_{\rm b}=8$.}
\label{fig:3_5}
\end{figure}

The extrapolated values  $\alpha_{\rm c}(\Delta,\Lambda\to 1)$ for the Ohmic
case are summarized in Fig.~\ref{fig:3_5}. In the limit of small $\Delta$, the
NRG result is in good agreement with the well established value
 $\alpha_{\rm c}(s=1,\Delta \to 0)=1$.
We estimate the error in  $\alpha_{\rm c}$ to be of
the order of $0.02$ which is due
to the various extrapolations just described. The solid line in
Fig.~\ref{fig:3_5} shows a linear fit to the numerical data
which gives $\alpha_{\rm c}(\Delta)=0.99 + 0.53\Delta$.
This is consistent with the RG result
$\alpha_{\rm c} = 1 + {\cal O}(\Delta/\omega_c)$ \cite{Leggett}.

\subsection{Scaling}
\label{sec:3C}

\begin{figure}[!t]
\epsfxsize=2.8in
\centerline{\epsffile{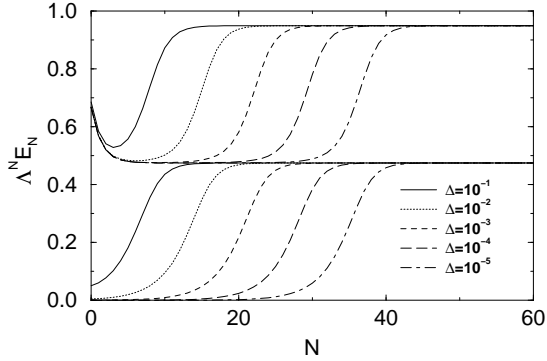}}
\caption{
Scaling of the flow of the many-particle levels $E_N(r)$ for
fixed $\alpha=0.6$, $s=1$, and various values of $\Delta$. The
NRG parameters are $N_{\rm s}=100$,
$N_{\rm b}=8$, and $\Lambda=2.0$.
}
\label{fig:3_6}
\end{figure}

We expect to observe scaling behavior in all physical properties
for fixed $\Delta$ and $\alpha\to\alpha_{\rm c}(\Delta)$ and for
fixed $\alpha$ and $\Delta\to 0$. Such a scaling can already be identified
on the level of the flow of the many-particle energies.
An example is shown in Fig.~\ref{fig:3_6} for fixed $\alpha=0.6$ and various
values of $\Delta$.
In this way we can easily determine the crossover scale
$T^\ast$ for the crossover from the localized to the delocalized fixed point
(there is only a single low-energy scale):
\begin{equation}
  T^\ast = {\rm const.}\times \Lambda^{-N^\ast} \ ,
\label{eq:TstarN}
\end{equation}
where we define $N^\ast$ as the value of $N$ where the first excited
state reaches the value $E_N = 0.3$. Note that a change of this
(arbitrary) value can be absorbed in a change of the prefactor in
eq.~(\ref{eq:TstarN}); this reflects the fact that a temperature
{\em scale} can only be defined up to a constant prefactor anyway.

\begin{figure}[!t]
\epsfxsize=2.8in
\centerline{\epsffile{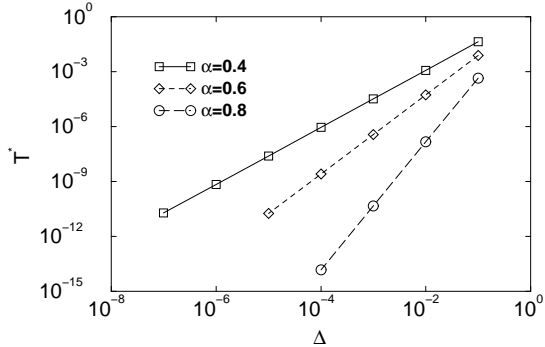}}
\caption{
Dependence of the crossover temperature $T^\ast$ on $\Delta$ for $s=1$ and
fixed values of $\alpha$. The exponents in $T^\ast \propto \Delta^x$
are $x\approx 1.55$ for $\alpha=0.4$, $x\approx 2.15$ for $\alpha=0.6$,
and $x\approx 3.49$ for $\alpha=0.8$.
 The
NRG parameters are $N_{\rm s}=100$,
$N_{\rm b}=8$, and $\Lambda=2.0$.
}
\label{fig:3_7}
\end{figure}

\begin{figure}[!t]
\epsfxsize=2.8in
\centerline{\epsffile{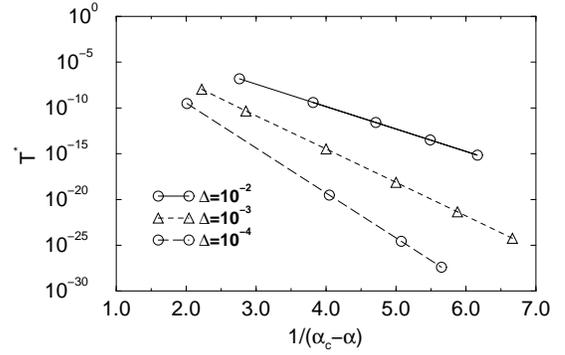}}
\caption[]{
Dependence of the crossover temperature $T^\ast$ on $\alpha$ for $s=1$
and fixed values of $\Delta$
(data for $\Delta=10^{-3}$ and $\Delta=10^{-4}$
same as in Fig.~4b of
Ref.~\onlinecite{BTV}).
The values for the critical coupling are $\alpha_{\rm c}=1.162$
for $\Delta=10^{-2}$, $\alpha_{\rm c}=1.150$ for
$\Delta=10^{-3}$, and $\alpha_{\rm c}=1.147$ for $\Delta=10^{-4}$. 
The NRG parameters are $N_{\rm s}=100$,
$N_{\rm b}=8$, and $\Lambda=2.0$.
}
\label{fig:3_8}
\end{figure}

In the scaling regime, the dependence of $T^\ast$ on
$\alpha$ and $\Delta$ is given by \cite{Leggett,expnote}
\begin{equation}
    T^\ast \propto \Delta^{1/(\alpha_{\rm c}-\alpha)} \ .
\label{eq:Tstar}
\end{equation}
As shown in Figs.~\ref{fig:3_7} and \ref{fig:3_8}, the NRG results
are in agreement with eq.~(\ref{eq:Tstar}).

\subsection{Flow for Sub-Ohmic Baths}
\label{sec:3D}

\begin{figure}[!t]
\epsfxsize=2.8in
\centerline{\epsffile{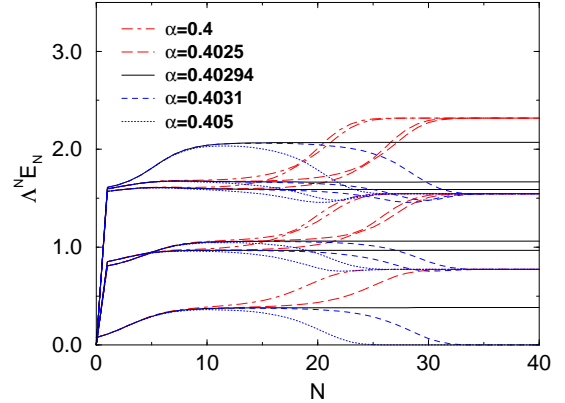}}
\caption{(Color online)
Flow diagram of the lowest lying many-particle energies
calculated with the star-NRG for the sub-Ohmic case ($s=0.8$,
$\Delta=0.1$),
using displaced oscillators as optimized basis.
The critical value is $\alpha_{\rm c}=0.40294$.
The NRG parameters are $N_{\rm s}=80$,
$N_{\rm b}=8$, and $\Lambda=2.0$.
}
\label{fig:flow_sub}
\end{figure}

As already mentioned in Sec.~\ref{sec:genstrat}, the chain-NRG
with a basis of undisplaced oscillators as in
eq.~(\ref{eq:basis-onex}) is sufficient
for the Ohmic and super-Ohmic case. Let us now turn to the
sub-Ohmic case where we expect problems with the chain-NRG
when the system is flowing to the localized fixed point.
Figure \ref{fig:flow_sub} shows a typical flow diagram of the
many-particle energies, calculated with the star-NRG for $s=0.8$
and a couple of $\alpha$ values close to the quantum critical
point $\alpha_{\rm c}=0.40294$.

In contrast to the Ohmic case, the transition in the sub-Ohmic case
is characterized by a {\em new} fixed point, the quantum critical fixed
point, with a level structure which is different from both the
localized and the delocalized fixed points. For any
$\alpha\ne\alpha_{\rm c}$ there is a finite crossover scale $T^\ast$
for the crossover to the localized fixed point
(for $\alpha>\alpha_{\rm c}$)
and to the delocalized fixed point (for $\alpha<\alpha_{\rm c}$).
The crossover scale can be defined in a similar way as in
Sec.~\ref{sec:3C}. A further analysis of the dependence of
 $T^\ast$ on $\vert\alpha-\alpha_{\rm c}\vert$ gives the
critical exponents. Their $s$-dependence has been shown already
in Fig.~5a of Ref.~\onlinecite{BTV}. A detailed investigation of the
critical properties of the sub-Ohmic spin-boson model will
appear elsewhere.

Here we focus on the level structure of the localized and delocalized
fixed point in Fig.~\ref{fig:flow_sub}. Both fixed points have exactly
the same level structure apart from an additional two-fold degeneracy
of all levels of the localized fixed point. This is evident from
Fig.~\ref{fig:flow_sub} (levels for $\alpha>\alpha_{\rm c}$
and $\alpha<\alpha_{\rm c}$ converge to the same spectrum)
 and also follows from the discussion
of Sec.~\ref{sec:oblfp}. However, the proper description of the
localized fixed point can only be achieved using an optimized basis
with displacements calculated as discussed in Sec.~\ref{sec:genstrat}
and Appendix \ref{app:B}.
Using a basis of undisplaced oscillators ($\theta=0$) leads to an incorrect
level structure. This can be seen in the upper right-hand panel of
Fig.~3 in Ref.~\onlinecite{BTV} ($s=0.6$, $\alpha>\alpha_{\rm c}$)
where the basis (\ref{eq:basis-onex}) was used. The resulting fixed point
levels are therefore not the same as the one for $\alpha<\alpha_{\rm c}$
in the upper left-hand panel of Fig.~3 in Ref.~\onlinecite{BTV}.

As mentioned in Sec.~\ref{sec:genstrat},
we did not yet succeed to implement the
concept of displaced oscillators in the chain-NRG, so the proper description
of the localized fixed point for $s<1$ is presently only possible with the
star-NRG. Fortunately, the problems of the chain-NRG
only show up when the flow is approaching the localized fixed point.
We can therefore
safely extract all the critical properties such as critical exponents
from the chain-NRG, as has been done in Ref.~\onlinecite{BTV}.

\begin{figure}[!t]
\epsfxsize=2.8in
\centerline{\epsffile{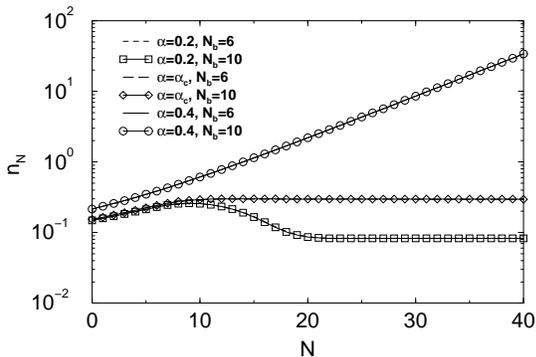}}
\caption{
Dependence of the expectation value
$n_{N} = \langle
b_{N}^\dagger b_{N} \rangle$ on the iteration number
$N$ for $s=0.8$, $\Delta=0.01$,
$\alpha=0.2<\alpha_{\rm c}$ (squares),
$\alpha=\alpha_{\rm c}=0.21488785$ (diamonds), and
$\alpha=0.4>\alpha_{\rm c}$ (circles),
and two different values of $N_{\rm b}$, calculated with the
star-NRG.
}
\label{fig:n_divergence}
\end{figure}

On the other hand, the use of a basis of displaced oscillators
within the star-NRG solves the problem of the boson-number divergence
(see Sec.~\ref{sec:genstrat}). This is illustrated in
Fig.~\ref{fig:n_divergence} where the dependence of the expectation value
$n_{N} = \langle b_{N}^\dagger b_{N} \rangle$ is shown for three
values of $\alpha$ ($\alpha=0.2<\alpha_{\rm c}$,
$\alpha=\alpha_{\rm c}=0.21488785$, and
$\alpha=0.4>\alpha_{\rm c}$) and two values of $N_{\rm b}$.
For all values of $\alpha$ we observe a rapid convergence
with $N_{\rm b}$, similar to the convergence shown for
$\alpha<\alpha_{\rm c}$ and $\alpha=\alpha_{\rm c}$ in
Fig.~\ref{fig:nb}. The difference here is that the data converge
with $N_{\rm b}$ also for $\alpha>\alpha_{\rm c}$ which can
{\em not} be achieved by using the
basis (\ref{eq:basis-one}), see Fig.~\ref{fig:nb}. Furthermore,
the expectation value $n_{N}$ diverges exponentially with $N$
for $\alpha>\alpha_{\rm c}$, as expected from the discussion
in  Sec.~\ref{sec:oblfp}. A diverging boson number itself is therefore
not a problem for the bosonic NRG, provided a proper optimized basis
is chosen.

Finally, a few words on the limitations of the star-NRG. Whereas
the localized fixed point is described correctly, the star-NRG
seems to fail in other respects:
the low-energy flow to the delocalized fixed point appears incorrect,
and critical exponents of the quantum phase transition deviate
from the chain-NRG results (and from analytically known values).
We do not yet fully understand this problem,
but it might be connected to truncation errors which affect the star-NRG
in a completely different way as the chain-NRG.
(The idea is that the truncation somehow
affects the character of the impurity operator to which the
added bosonic site couples in each step.)
The precise characterization of this problem and its possible solution are
left for future studies.

\section{Thermodynamic Quantities}
\label{sec:thdyn}
In this section, we describe how thermodynamic quantities
can be extracted from the flow of many-particle levels
$E_N(r)$ which are calculated with the bosonic NRG.
Starting from the $E_N(r)$ there is no difference
(from a technical point of view) between the fermionic
and the bosonic case (for the fermionic case see, for example,
Refs.~\onlinecite{Kri80,Oli94}).
Nevertheless, for completeness we include a brief discussion
of the technical details here. We show results for the
impurity contribution to the entropy and the specific heat
in the Ohmic case (using the chain-NRG with basis (\ref{eq:basis-onex})).
The Ohmic
case has been studied in detail in Refs.~\onlinecite{Costi98,CZ}
(for earlier work on thermodynamic properties see
Refs.~\onlinecite{Leggett,Goehrlich88,Sassetti90}).
The agreement with the results from Refs.~\onlinecite{Costi98,CZ} is
excellent
which again confirms the reliability of the bosonic NRG for the
investigation of quantum impurity models involving a bosonic
bath. A few comments on thermodynamic properties in the
sub-Ohmic case are given at the end of this section.

Consider the spectrum of many-particle energies $E_i$ of a
discretized version of the spin-boson model (not necessarily
the discretized Hamiltonians (\ref{eq:hstar}) and (\ref{eq:hchain})
used in the bosonic NRG).
The grand canonical partition function,
$Z = {\rm Tr}\,e^{-\beta(H-\mu N )}$,
reduces to
\begin{equation}
   Z = \sum_i e^{-\beta E_i} \ ,
\label{eq:ZE}
\end{equation}
as the chemical potential $\mu$ is set to zero (we are
interested in gapless spectral functions $J(\omega)$).
Free energy $F$ and entropy $S$ are then given by
\begin{equation}
   F = - T \ln Z  \ \ {\rm and}
  \ \
  S= -\frac{\partial F}{\partial T} \ .
\end{equation}
(We set $k_{\rm B}=1$.)
The impurity contribution to
the entropy is
\begin{equation}
  S_{\rm imp} = S - S_0 \ ,
\end{equation}
where $S$ is the entropy of the full system and $S_0$
the entropy of the system without impurity.

Before we discuss the full temperature dependence of
$S_{\rm imp}(T)$, let us focus on the value of
$S_{\rm imp}$ at the localized and delocalized fixed points:
$S_{\rm imp,L}$ and $S_{\rm imp,D}$. It is well known
that $S_{\rm imp,L}=\ln 2$ and $S_{\rm imp,D}=0$\cite{Costi98,CZ},
but it might not be obvious that these values can be directly extracted
from the many-particle spectra at the fixed points.

In Sec.~\ref{sec:3A} we already showed that the fixed point spectrum
of the delocalized fixed point is the same as the one of a free bosonic
chain, which is nothing else but the system without impurity.
This implies that for the delocalized fixed point
\begin{equation}
E_i = E_{i,0} + \Delta E \ ,
\label{eq:Ei}
\end{equation}
 with
$E_i$ ($E_{i,0}$) the many-particle energies of the system
with (without) impurity and $\Delta E$ a constant shift independent
of $i$. It is clear that this equation cannot hold for {\em all}
levels, it is only valid for energies sufficiently below the
crossover scale to the fixed point.

Equation (\ref{eq:Ei}) directly leads to the proof of  $S_{\rm imp,D}=0$:
we have $Z_{\rm D}=\exp[-\beta\Delta E] Z_{\rm 0}$, and from this
$F_{\rm D} = F_{\rm 0}+\Delta E$.
The energy shift drops out in the derivative so that $S_{\rm D}=S_0$ and
the impurity contribution to the entropy at the delocalized
fixed point is given by $S_{\rm imp,D}=0$.

In a similar way one can easily prove that
$S_{\rm imp,L}=\ln 2$: in this case we have
\begin{equation}
   Z_{\rm L} = 2 \sum_i e^{-\beta E_i} \ , \ \
 E_i = E_{i,0} + \Delta E\ ,
\end{equation}
with the factor of 2 due to the additional
double degeneracy of all many-particle levels at the
localized fixed point. This gives
$Z_{\rm L}=2 \exp[-\beta\Delta E] Z_{\rm 0}$,
and from this
$F_{\rm L} = -T \ln 2 + F_{\rm 0}+\Delta E$ and $S_{\rm L}=\ln 2 + S_0$, corresponding
to $S_{\rm imp,L}=\ln 2$.
From this discussion it follows
that $S_{\rm imp,L}=\ln 2$ and $S_{\rm imp,D}=0$
independent of the exponent $s$ in the spectral function $J(\omega)$.

For any finite $\Delta$ and $\alpha$,
the values $S_{\rm imp,L}=\ln 2$ and $S_{\rm imp,D}=0$
are strictly valid only in the limit $T\to 0$. Note that
a proper definition of these zero-point entropies requires
the correct order of limits:
the thermodynamic limit has to be taken {\em before} the
limit $T\to 0$. With the order of the limits reversed,
the zero-point entropy would be equal to $\ln d_{\rm g}$,
with $d_{\rm g}$ the degeneracy of the ground state.
Although this happens to give the same values for
$S_{\rm imp,L}$ and $S_{\rm imp,D}$ in the case studied here,
this equivalence is not generally valid.
(This can be seen, for example, in the NRG calculations for the
single-impurity Anderson model \cite{Kri80}
where the degeneracy of the ground state oscillates between
1 for even and 4 for odd iterations when the system approaches
the fixed point of a screened spin, which has $S_{\rm imp} = 0$.
Also, any non-trivial quantum critical fixed point is expected
to have a residual entropy which is not $\ln d_{\rm g}$ with
integer $d_{\rm g}$.)

The impurity contribution to the entropy is close to
a fixed point value also when the system is close to this
fixed point in an intermediate range of the flow diagram.
From Fig.~\ref{fig:3_1}a we can therefore immediately see that the temperature
dependence of $S_{\rm imp}(T)$ contains a crossover from
a high-temperature value $S_{\rm imp}\approx S_{\rm imp,L} =\ln 2$ to the
low-temperature value $S_{\rm imp}(T\to 0)=S_{\rm imp,D}=0$;
provided the flow is to the delocalized fixed point.
The detailed behavior of $S_{\rm imp}(T)$ in the crossover
region requires a numerical calculation as described below.

In the bosonic NRG, we do not have access to the full spectrum
of many-particle energies $E_i$ as used in eq.~(\ref{eq:ZE}).
Instead, the iterative procedure results in a sequence of
many-particle energies $E_N(r)$ with iteration number $N$
and $r=1,\ldots N_{\rm s}$. According to the discussion in
Refs.~\onlinecite{Wil75,Kri80}, each of the sets of many-particle
energies is assumed to be a good description of the system
for a certain temperature $T_N$ with
\begin{equation}
   T_N = x \omega_c \Lambda^{-N} \ ,
\label{eq:TN}
\end{equation}
with $x$ a dimensionless constant of the order of $1$,
chosen such that $T_N$ lies within the spectrum
$E_N(r)$.

For each iteration step $N$, the partition function is
calculated for the temperature $T_N$:
\begin{equation}
   Z_N = \sum_r e^{- E_N(r)/T_N} \ .
\end{equation}
In addition, the internal energy at iteration step $N$
for the temperature $T_N$ is calculated as
\begin{equation}
   E_N = \frac{1}{Z_N}\sum_r E_N(r) e^{- E_N(r)/T_N} \ .
\end{equation}
This is the information we have available for the numerical
calculation of thermodynamic properties.

One possibility to proceed is to calculate the free energy
$F_N = - T_N \ln Z_N$ for each iteration step, and
from this the entropy  $S= -\frac{\partial F}{\partial T}$
via a discrete differentiation. This procedure has been shown
to give good results in the fermionic case (see, for example,
Ref.~\onlinecite{BH}). It requires, however, a precise calculation of
the difference of the ground state energies between subsequent
steps; this appears to introduce some errors in the calculations
within the bosonic NRG. (In general, the bosonic NRG is less
accurate in the calculation of thermodynamic properties as
compared to the fermionic NRG because we cannot keep
as many states as in  the fermionic case.)

 Therefore, we use an alternative
approach in which the entropy $S_N$ at iteration step $N$
for the temperature $T_N$ is calculated via
\begin{equation}
   S_N = \frac{E_N}{T_N} +  \ln Z_N \ .
\end{equation}
This approach avoids the discrete differentiation, and
does not require the knowledge of the ground state energies.

\begin{figure}[!t]
\epsfxsize=3.0in
\centerline{\epsffile{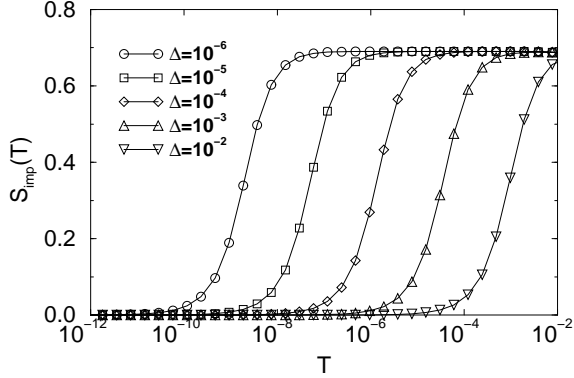}}
\caption{Temperature dependence of the impurity contribution
to the entropy, $S_{\rm imp}(T)$, for $\alpha = 1/3$, $s=1$ (Ohmic case),
and various values of $\Delta$.
}
\label{fig:4_1}
\end{figure}

Let us now discuss the results for entropy and specific heat
calculated with the bosonic NRG using the method
just described. Figure \ref{fig:4_1} shows the temperature
dependence of the impurity contribution to the
entropy, $S_{\rm imp}(T)$,  for $\alpha = 1/3$, $s=1$ (Ohmic case),
and various values of $\Delta$. We observe a crossover
from the high-temperature value $S_{\rm imp}=\ln 2$ to the
low-temperature value $S_{\rm imp}=0$ at a crossover scale
$T^\ast$, which is the same as the one introduced in
Sec.~\ref{sec:3C}. The crossover scale decreases with
decreasing $\Delta$
in agreement with eq.~(\ref{eq:Tstar}). Note the similarity
of Fig.~\ref{fig:4_1} to Fig.~\ref{fig:3_6} for the scaling
of the energy levels, a similarity which is simply due to
the relation between $S_{\rm imp}(T)$ and the flow of the
many-particle levels.

\begin{figure}[!t]
\epsfxsize=3.0in
\centerline{\epsffile{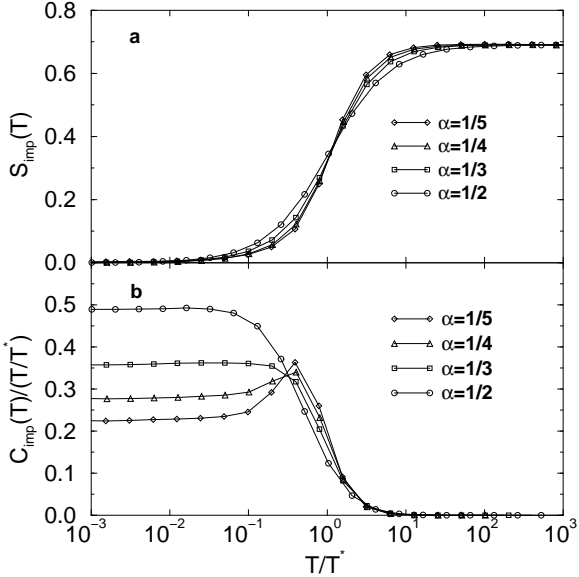}}
\caption{(a) Scaling curves  of the impurity contribution
to the entropy, $S_{\rm imp}(T)$, for  $s=1$ (Ohmic case),
and various values of $\alpha$;
(b) Scaling curves  of the impurity contribution
to the specific heat, $C_{\rm imp}(T)/(T/T^\ast)$, for the same
parameters as in (a).
}
\label{fig:4_2}
\end{figure}

As briefly mentioned in Sec.~\ref{sec:3A}, the vicinity to the
localized fixed point for early iterations (which results
in the high-temperature value $S_{\rm imp}(T)\approx \ln 2$)
does not imply localization. The value of
$S_{\rm imp}(T)$ for high temperatures is due to the fact that for
temperatures $T\gg \Delta$ both states of the two-state system
contribute equally to the thermodynamics. Note also the
similarity to $S_{\rm imp}(T)$ in the Kondo model: there
the high-temperature phase is that of a local moment with
both spin $\uparrow$ and  $\downarrow$ configurations
contributing to the entropy (a temperature dependence
of $S_{\rm imp}(T)$ as in Fig.~\ref{fig:4_1} might therefore
appear more natural in the Kondo model but, of course, it
is also valid here).

The scaling behavior of $S_{\rm imp}(T)$ for fixed $\alpha = 1/3$
and various $\Delta$ is obvious and is shown in Fig.~\ref{fig:4_2}a
together with the scaling curves for $\alpha =1/5$, $1/4$, and $1/2$.
The agreement with the exact results from the Bethe Ansatz calculations
in Ref.~\onlinecite{CZ} is very good (see Fig.~7a in Ref.~\onlinecite{CZ}), in
particular for the $\alpha$-dependence of the scaling curves.

The temperature dependence of the specific heat,  $C_{\rm imp}(T)$,
is calculated via $C_{\rm imp}(T)/T = \partial S_{\rm imp}(T)/\partial T$.
Here we cannot avoid the discrete differentiation of $S_{\rm imp}(T)$.
The scaling of $S_{\rm imp}(T)$ implies a scaling of $C_{\rm imp}(T)/T$
as shown in Fig.~\ref{fig:4_2}b. This figure is also very similar to
previous calculations (see Fig.~2 in Ref.~\onlinecite{Costi98} from the
NRG via mapping to the anisotropic Kondo model and
Fig.~7b in Ref.~\onlinecite{CZ} using the Bethe Ansatz), and we find the
same characteristic features here: a linear specific heat $C\propto T$
for low temperatures, a peak in $C/T$ at $T\approx T^\ast$ for
small dissipation $\alpha<0.3$ in contrast to the monotonous decrease
of  $C/T$ for large dissipation $\alpha>0.3$, and a characteristic
crossing point of all the $C/T$ scaling curves.

Similar to the NRG calculations in Ref.~\onlinecite{Costi98}, the thermodynamic
quantities can only be calculated on a discrete mesh of temperatures
given by eq.~(\ref{eq:TN}). This strongly limits the resolution
of the peak in $C/T$ for $\alpha<0.3$, in contrast to the Bethe Ansatz
calculations of Ref.~\onlinecite{CZ}.

The physics in the sub-Ohmic case is much richer, due to the
appearance of a line of quantum critical points \cite{BTV}.
This is reflected in the behavior of the entropy and the
specific heat. For the results of $S_{\rm imp}(T)$
and $C_{\rm imp}(T)$ close to the quantum critical points
we refer to a subsequent publication.
Here we focus on the flow  to the delocalized phase.

\begin{figure}[!t]
\epsfxsize=3.0in
\centerline{\epsffile{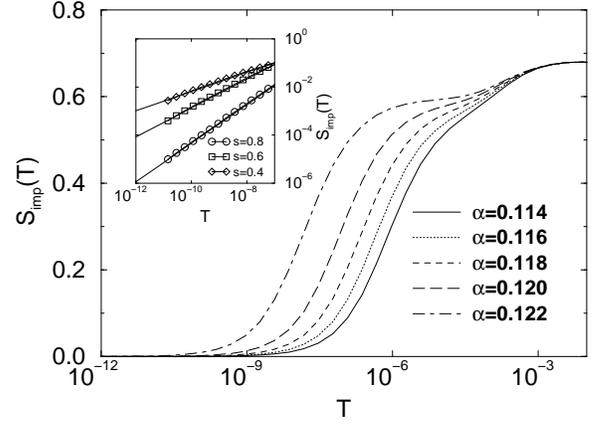}}
\caption{Temperature dependence of the impurity contribution
to the entropy, $S_{\rm imp}(T)$, in the sub-Ohmic case
for various values of $\alpha$ and  $s=0.8$ (main panel)
and various values of $s$ (inset). The coupling $\alpha$
is below $\alpha_{\rm c}$ so that the flow is to the delocalized
phase for all parameters in this figure. Lines with symbols in the
inset are data from the bosonic NRG and solid lines are fits
assuming a power-law, $S_{\rm imp}(T)\propto T^s$.
}
\label{fig:4_3}
\end{figure}

Figure \ref{fig:4_3} shows the temperature dependence of the
impurity contribution
to the entropy, $S_{\rm imp}(T)$, in the sub-Ohmic case, $s=0.8$,
for various values of $\alpha$  below the
critical value $\alpha_{\rm c}\approx 0.125$. For $\alpha$ close to
$\alpha_{\rm c}$ we observe a two stage quenching of the
entropy of the free moment (the quantum critical point has
a non-trivial zero-point entropy of $S_{\rm qcp}(T\to0)\approx 0.6$
for $s=0.8$). As expected, the temperature scale for the crossover
to the delocalized fixed point increases with the distance
from the critical point. The low-temperature behavior of $S_{\rm imp}(T)$
for $\alpha<\alpha_{\rm c}$ is given by $S_{\rm imp}(T)\propto T^s$
which can be seen more clearly in the inset of
Fig.~\ref{fig:4_3} where  $S_{\rm imp}(T)$ is plotted for various
values of $s$.
This behavior is in agreement with the calculations of
Ref.~\onlinecite{Goehrlich88}, where $C(T)\propto T^s$
was found for the slightly asymmetric ($\epsilon\ne0$)
sub-Ohmic spin-boson model.
[While the finite $\epsilon$ turns the quantum phase transition into a smooth crossover,
it does not influence the qualitative low-energy behavior in the delocalized phase.]

The data in Fig.~\ref{fig:4_3} are calculated with the
chain-NRG. The results from the star-NRG look similar (they
give, in particular the correct value $S_{\rm imp}(T\to0)=\ln 2$
if the flow is to the localized phase). We observe, however, a
low-temperature behavior for  $S_{\rm imp}(T)$ which is
{\em different}
from the correct form, $S_{\rm imp}(T)\propto T^s$. As briefly
mentioned in Sec.~\ref{sec:3D}, the reason for this failure of the star-NRG
is presently not clear but probably due to truncation errors.

Despite these deficiencies,
the bosonic NRG is a reliable tool for the calculation of thermodynamic
properties in a wide range of parameters and the comparison with
well-established results is very promising.
Thermodynamic properties in the
quantum critical region will be discussed in a separate publication.

\section{Dynamic Quantities}
\label{sec:dyn}
The calculation of dynamic properties is straightforward
within the bosonic NRG and proceeds in a very similar way
(from a technical point of view) as in the fermionic case.
The typical problems such as the combination
of information from different iteration steps and the broadening of the
discrete spectra have been discussed already in the literature
(see, for example, Refs.~\onlinecite{Fro86,Sak89,Cos92,Cos94,BCV})
and need not be repeated here.

\subsection{Dynamical Spin Correlations}

One important dynamic quantity of interest in the spin-boson model
is the spin-spin correlation function (spin
autocorrelation function)
\begin{equation}
C(\omega)=\frac{1}{2\pi} \int_{-\infty}^{+\infty} e^{i \omega t} C(t)\,
{\rm d}t
\ ,
\end{equation}
with $C(t) = \frac{1}{2}\langle [\sigma_z(t),\sigma_z]_+ \rangle$.
We only consider {\em equilibrium} correlation functions,
in general for finite temperatures, but the focus here is on $T=0$
so that the expectation value $\langle \ldots \rangle$ has to
be taken with respect to the ground state.

For a discrete Hamiltonian, the spin-spin correlation function
at $T=0$ can be written
as
\begin{equation}
     C(\omega)=\frac{1}{2}\sum_{n} | \langle 0 | \sigma_z | n \rangle |^2
               \delta \left( \omega+\epsilon_0-\epsilon_n \right)
      \,\,\,\,\, , \  \omega > 0,
\label{eq:Cdisc}
\end{equation}
with $C(\omega)=C(-\omega)$. 
Note that with the above definition of $C(t)$, the quantity
$C(\omega)$ is purely real and related to the imaginary
part of the spin-susceptibility $\chi(\omega)$ via
$C(\omega)= \frac{1}{2}\pi \vert {\rm Im}\chi(\omega)\vert$ (see also
eq.~(3.96) in Ref.~\onlinecite{Leggett}). 

\begin{figure}[!t]
\epsfxsize=3.0in
\centerline{\epsffile{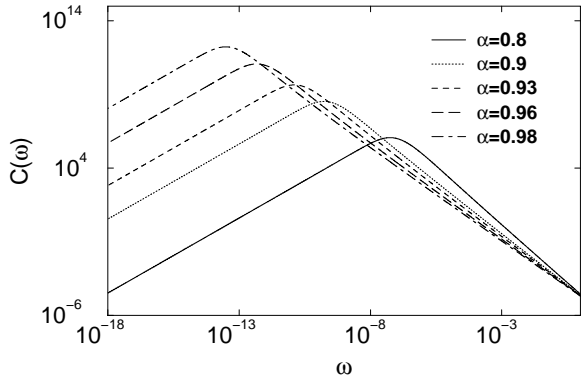}}
\caption{
Spin-spin correlation function $C(\omega)$ calculated for
the Ohmic case, $s=1$, $\Delta=0.01$,
and various values of $\alpha<\alpha_{\rm c}$ close to the transition.
For small frequencies, $C(\omega)\propto \omega$, whereas for
higher frequencies we observe a divergence, $C(\omega)\propto \omega^{-1}$,
with logarithmic corrections.
}
\label{fig:5_1}
\end{figure}

Due to the truncation in the course
of the iterative diagonalization, we cannot calculate $C(\omega)$
simultaneously for all energy scales. Instead, the correlation function
is calculated for each cluster of length $N$ (which gives information
on energy scales of the order of $\Lambda^{-N}$) and this information
has to be added up properly. Finally, the
discrete spectrum has to be broadened which results in continuous curves for
$C(\omega)$ as shown, for example, in Fig.~\ref{fig:5_1}.

These technical issues are dealt with using the approach described
in Ref.~\onlinecite{BCV}; to broaden the spectra, we use a Gaussian
on a logarithmic scale (see eq.~(8) in Ref.~\onlinecite{BCV}) with
broadening parameter $b=0.7$.

We also define the correlation function $S(\omega)$ as
\begin{equation}
     S(\omega)=\frac{2C(\omega)}{\omega^{s}} \, .
\end{equation}
The static spin-susceptibility
$\chi$ is defined as
\begin{equation}
 \chi=2 \frac{\partial\langle \sigma_z\rangle}{\partial \epsilon }
\ \ , \ \  \langle \sigma_z\rangle = \langle 0\vert\sigma_z \vert 0\rangle \, .
\end{equation}
It is related to $C(\omega)$ via
\begin{equation}
  \chi=4 \int_{0}^{\infty} \frac{C(\omega)}{\omega} \, {\rm d}\omega \ ,
\end{equation}
and, using eq.~(\ref{eq:Cdisc}), can be written in the form
\begin{equation}
 \chi=2 \sum_{n} \frac{ |\langle 0 | \sigma_z | n\rangle |^2}{\epsilon_n -\epsilon_0} \, .
 \label{eq:chi_disc}
\end{equation}
Here, we calculate the susceptibility according to eq.~(\ref{eq:chi_disc}).

The bosonic NRG allows the calculation of dynamic properties in a
wide range of frequencies so that the functional dependence of
$C$ on the frequency $\omega$ (such as a power-law
$C(\omega)\propto\omega^s$) can be easily extracted.
 However, the exponent
of the calculated $C(\omega)$ has a deviation from the expected
value $s$ (for the flow to the delocalized fixed point) of about $2\%$.
To extract the correct prefactor of $C(\omega)$ (which we use to
compare with the exact results at the Toulouse
point and to check the Shiba relation),
we need to redefine the quantity $S(\omega)$ as
\begin{equation}
     S(\omega)=\frac{2C(\omega)}{\omega^{\delta}} \, ,
\label{eq:Sdelta}
\end{equation}
where $\delta$ is the exponent fitted to  $C(\omega)$ in the small
frequency regime.

In Fig.~\ref{fig:5_1}, $C(\omega)$ is shown for the Ohmic case
and  a set of $\alpha$ values
close to the critical $\alpha_{\rm c}$.
The spin-spin correlation function shows
the expected power-law behavior, $C(\omega)\propto \omega$,
in the low-frequency regime $\omega < T^{\ast}$.
In the limit of $\alpha \rightarrow \alpha_{\rm c}$,  $C(\omega)$ shows
a divergence for $\omega > T^{\ast}$, $C(\omega)\propto \omega^{-1}$,
with logarithmic corrections.


\begin{figure}[!t]
\epsfxsize=3.0in
\centerline{\epsffile{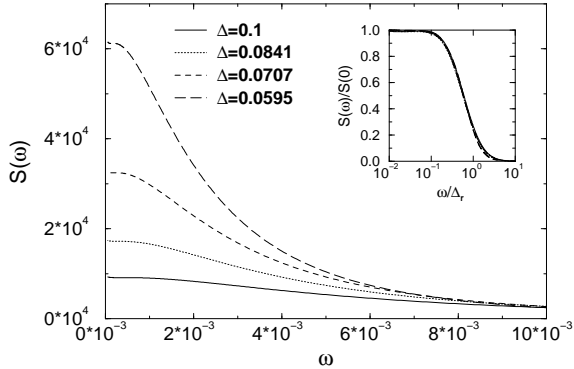}}
\caption{
Spin-spin correlation function $S(\omega)=2C(\omega)/\omega^\delta$
at the Toulouse point $\alpha=\frac{1}{2}$ for the Ohmic case $s=1$ and
various values of $\Delta$. The inset shows the scaling of these
curves ($S(\omega)/S(0)$ plotted versus $\omega/\Delta_{\rm r}$) together
with the exact result (thick dashed line).
}
\label{fig:5_2}
\end{figure}

In Fig.~\ref{fig:5_2}, $S(\omega)$ is plotted at the
Toulouse point ($\alpha=\frac{1}{2}$) of  the Ohmic spin-boson model
for several values of $\Delta$. At this point, the Ohmic
spin-boson model is exactly solvable, as discussed in
Ref.~\onlinecite{Weiss}.
In the inset of Fig.~\ref{fig:5_2}, all the curves
are rescaled onto one  curve with a renormalized tunneling amplitude
$\Delta_{\rm r}$.
Here, $\Delta_{\rm r}$ is defined as
\begin{equation}
\Delta_{\rm r,NRG}= \frac{\chi_{\rm e} \Delta_{\rm r,e}}{\chi_{\rm NRG}}
 \, ,
\end{equation}
where $\chi_{\rm e}$ is the exact susceptibility and $\Delta_{\rm r,e}$
the exact renormalized tunnelling amplitude
at the Toulouse point: $\Delta_{\rm r,e}=\pi \Delta^2 /2$
and $\chi_{\rm e} \Delta_{\rm r,e}=8/\pi$.
The quantity  $\chi_{\rm NRG}$
is the susceptibility calculated from the NRG, eq.~(\ref{eq:chi_disc}).
The comparison of the result from the bosonic NRG
with the exact rescaled $S(\omega)/S(0)$
 shows good
agreement (see the inset of Fig.~\ref{fig:5_2}).
The exact result is given by \cite{Weiss}:
\begin{equation}
    \frac{S(\omega)}{S(0)}=\frac{1}{8\left(1+x^2 \right)} 
\left[\frac{\ln \left(1+4x^2\right)}{x^2} + 
             \frac{2 {\rm arctan} \left( 2x \right)}{x}\right]    \, ,
\end{equation}
with $x=\frac{\omega}{\Delta_{\rm r,e}}$.

\begin{figure}[!t]
\epsfxsize=3.0in
\centerline{\epsffile{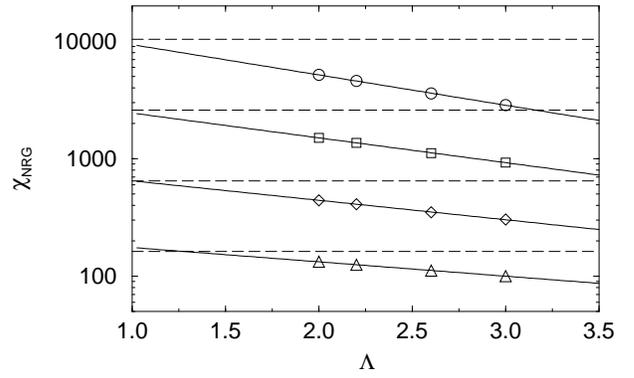}}
\caption{NRG results for $\chi_{\rm NRG}$ calculated
at the Toulouse point $\alpha=\frac{1}{2}$ 
for the Ohmic case $s=1$ and various values of $\Lambda$
and $\Delta$ (circles: $\Delta=0.0125$, squares: $\Delta=0.025$,
diamonds: $\Delta=0.05$, triangles: $\Delta=0.1$). Dashed
lines show the exact values 
$\chi_{\rm e}=16/ \left(\pi^2 \Delta^2 \right)$ and thin
solid lines are fits to the numerical results.
}
\label{fig:5_new}
\end{figure}

The NRG results for $\chi_{\rm NRG}$ deviate significantly from the
exact value $\chi_{\rm e}=16/ \left(\pi^2 \Delta^2 \right)$. 
However, as shown in Fig. \ref{fig:5_new},
this deviation is entirely due to discretization effects
and the extrapolation $\Lambda\to 1$ shows almost perfect
agreement with the exact result. Note that the exact value for
$\chi_{\rm e}$ has been obtained for a {\em soft} cut-off
in the bath spectral function,
$J(\omega)= 2 \pi \alpha \omega \exp(-\omega/\omega_{\rm c})$. 
To allow for a comparison,
the logarithmic discretization has to be performed for
the same soft cut-off  
(we introduce a high-energy hard
cut-off at $\omega = 15 \omega_{\rm c}$).

\begin{figure}[!t]
\epsfxsize=3.0in
\centerline{\epsffile{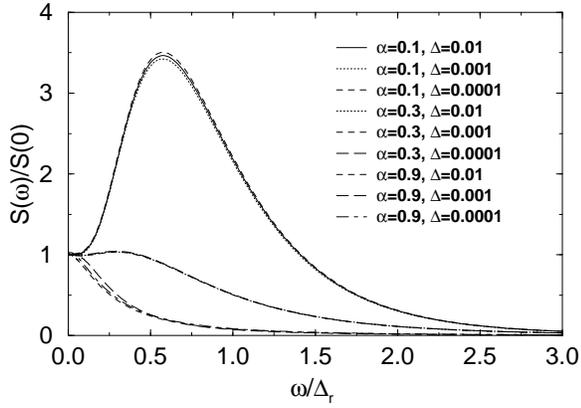}}
\caption{Scaling spectra for the spin-spin correlation function
$S(\omega)=2C(\omega)/\omega^\delta$ for various values of
 $\alpha$ in the Ohmic case $s=1$.
}
\label{fig:5_3}
\end{figure}

The scaling behavior of $S(\omega)/S(0)$ for fixed $\alpha$ and
different values of $\Delta$ is shown in Fig.~\ref{fig:5_3}.
For this we need to identify an energy scale $T^\ast$ as in Sec.~\ref{sec:3C}.
There are, as usual, various possibilities to define the energy
scale: the position of the peak in $C(\omega)$, $\omega_{\rm p}$,
the quantity $1/\chi$, and the $T^\ast$ as defined in
eq.~(\ref{eq:TstarN}).
Obviously, we have
$T^{\ast} \propto \omega_{\rm p} \propto 1/\chi \propto \Delta_{\rm r}$,
and we choose $\Delta_{\rm r} = 8/(\pi \chi)$
for the energy scale in Fig.~\ref{fig:5_3}.
The scaling curves shown in Fig.~\ref{fig:5_3} are in good agreement
with the ones calculated in Ref.~\onlinecite{Costi96};
in particular, we find that
the coherent peak in $S(\omega)$ disappears  when
 $\alpha$ is larger than $\alpha^{*} \approx 0.3$.

In our notation, the Shiba relation reads\cite{Sassetti90}
\begin{equation}
   2\alpha \left( \frac{\chi}{2} \right)^2= S(0) \,.
\end{equation}
Table \ref{tab:shiba} shows the results from the bosonic NRG for the Ohmic
case and various values of $\alpha$ and $\Delta$. The parameter
$\delta$ is the exponent defined in eq.~(\ref{eq:Sdelta}). We find that
the Shiba relation is fulfilled within an error of about  $10 \%$.


\protect\begin{table}
\caption{
Results from the bosonic NRG for the Shiba-relation in the Ohmic case
for  various values of $\alpha$ and $\Delta$.
}
\label{table1}
\begin{tabular}{ccccccc}
$s$
& $\alpha$
& $\Delta$
& $\delta$
& $2\alpha \left( \frac{\chi}{2} \right)^2$
& $S(0)$
& $\%$error \\
$1.0$ &   $0.02$ &    $0.005$ &   $1.018$  &  $0.201\times 10^4$ &  $0.221 \times 10^4$  &  $9.9$$\%$  \\
$1.0$ &   $0.1$ &     $0.01$  &   $1.018$  &  $0.603\times 10^4$ &  $0.631 \times 10^4$  &  $4.6$$\%$  \\
$1.0$ &   $0.4$ &     $0.01$  &   $1.018$  &  $0.294\times 10^7$ &  $0.308 \times 10^7$  &  $4.9\%$  \\
$1.0$ &   $0.5$ &     $0.025$ &   $1.018$  &  $0.154\times 10^7$ &  $0.163 \times 10^7$  &  $5.8\%$  \\
$1.0$ &   $0.7$ &     $0.03$  &   $1.019$  &  $0.376\times 10^9$ &  $0.416 \times 10^9$  &  $10.6\%$ \\
$1.0$ &   $0.9$ &     $0.1$   &   $1.018$  &  $0.172\times 10^{10}$ & $0.192 \times 10^{10}$ &  $11.6\%$ \\
\end{tabular}
\label{tab:shiba}
\end{table}

\subsection{Order Parameter}

In the localized phase, which corresponds to the ordered phase
of the $1/r^2$ Ising model, it is straightforward to define an
order parameter, $m$, corresponding to the magnetization of the Ising
model.
In the language of the spin-boson model, it corresponds to the
static, i.e., $\omega=0$, part of the spin autocorrelation function;
in the language of the anisotropic Kondo model this just measures the
prefactor of the Curie part of the local susceptibility, i.e.,
the unscreened fraction of the impurity moment.
The Kosterlitz-Thouless nature of the transition implies a {\em jump}
of the order parameter at the phase transition.

\begin{figure}[!t]
\epsfxsize=3.0in
\centerline{\epsffile{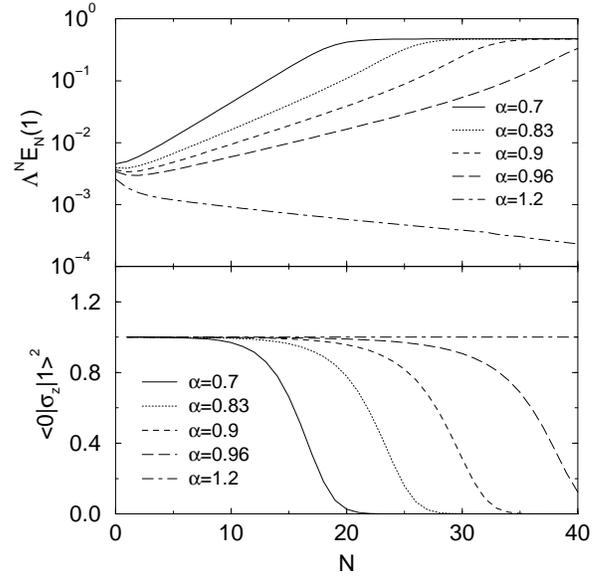}}
\caption{(a) Energy of the first excited state $E(1)$ versus iteration
number for $s=1.0$, $\Delta=0.01$, and various values of $\alpha$.
(b) matrix element $\vert\langle 0 | \sigma_z | 1\rangle\vert^2$
versus iteration number for the same parameters as in (a).
}
\label{fig:5_4}
\end{figure}

We extract this order parameter from the $\delta(\omega)$ contribution
to $C(\omega)$. The identification of such a $\delta$-peak in the
spectrum of $C(\omega)$ requires some extra care.
The spectrum calculated with the NRG consists of $\delta$-peaks only, which
have to be broadened suitably to give spectra as shown, for example,
in Fig.~\ref{fig:5_1}. Therefore, one has to decide whether a
$\delta$-peak in the spectrum belongs to the continuum or
whether it survives as a $\delta$-peak in the thermodynamic
limit \cite{delta-peak}.
The procedure is illustrated in Fig.~\ref{fig:5_4}. Let us first
note that the matrix element
$\vert\langle 0 | \sigma_z | 0\rangle\vert^2$
vanishes for all parameters of Fig.~\ref{fig:5_4}.
We therefore plot the matrix element
$\vert\langle 0 | \sigma_z | 1\rangle\vert^2$ in
Fig.~\ref{fig:5_4}b together with the energy $E(1)$ in
Fig.~\ref{fig:5_4}a. We observe that for $\alpha>\alpha_{\rm c}$
the energy $E(1)$ vanishes {\em faster} than $\Lambda^{-N}$
with increasing iteration number $N$, whereas the
matrix element approaches a constant,
$\vert\langle 0 | \sigma_z | 1\rangle\vert^2 \to {\rm const} \approx 1$.
In the thermodynamic limit, this gives the $\delta$-peak
at $\omega=0$, with the weight given by the matrix element
 $\vert\langle 0 | \sigma_z | 1\rangle\vert^2$ which
corresponds to the order parameter $m$.
On the other hand, for $\alpha<\alpha_{\rm c}$ the energy $E(1)$
is proportional to $\Lambda^{-N}$, and the corresponding
$\delta$-peak is therefore interpreted as being part of the
continuum.

These arguments result in an order parameter $m(\alpha)$
which is zero for $\alpha<\alpha_{\rm c}$ and jumps to
a finite value for $\alpha\ge\alpha_{\rm c}$.
In the sub-Ohmic case, the order parameter shows power-law behavior
near the quantum phase transition, which will be discussed in detail
elsewhere.
As an aside, we note that the order parameter can also be extracted
from the Curie part of the static local susceptibility $\chi(T)$.

\section{Conclusions}
\label{sec:concl}
In this paper we have discussed a generalization of Wilson's NRG
technique to quantum impurity problems with a bosonic bath.
Focussing on the application to the spin-boson model,
we have shown that this novel method provides reliable
results for both static and dynamic quantities in the whole range
of model parameters and temperatures.
For the case of Ohmic damping, we have compared our data
to existing results in the literature and found good agreement.
The bosonic NRG is able to reproduce the expected scaling
behavior as function of temperature or frequency.
For sub-Ohmic damping, there is a line of continuous boundary
quantum phase transitions for all $0<s<1$, with exponents varying as
function of $s$; details of the associated quantum critical behavior
will be discussed in a forthcoming paper (see also Refs.~\onlinecite{BTV,VTB}).

We have outlined several details of the numerical implementation
of the bosonic NRG.
Two general strategies were discussed, termed chain-NRG and star-NRG:
both use a sequence of boson states with exponentially decreasing energy
scales, but in the chain-NRG the bath states form a chain and the impurity
couples to the first chain site only,
whereas in the star-NRG the impurity is coupled to all bath sites which are
not connected to each other.
The advantages and disadvantages of both methods were discussed in detail,
together with the important issue of the optimal choice of a basis
set of bosonic states at each bath site.
This problem is inherent to the bosonic NRG, as the infinite Hilbert space
has to be truncated, and specific solutions have to be found for the
problem at hand.
We have argued that in the Ohmic, super-Ohmic 
and sub-Ohmic cases  of the spin-boson model (except for the flow to the
localized fixed point in the latter case)
the basis formed by the lowest boson number eigenstates is sufficient,
and all fixed points are properly captured in the NRG.
Most of the results in this paper were obtained with this basis choice
using the chain-NRG method; we have given a detailed account
on convergence issues with respect to truncation and discretization
parameters.
In the sub-Ohmic case, the boson numbers diverge in the localized regime
in the low-energy limit, and a different basis choice is needed.
We have described suitable basis states using displaced harmonic
oscillators, which solve the problem for the star-NRG.
Open numerical issues include a reliable implementation of the displaced-oscillator
basis for the chain-NRG, a more accurate calculation of
dynamic quantities, as well as the numerical stability
for very long iterations, i.e.,
very small energy scales.

The bosonic NRG can be easily generalized to impurities with multiple
bosonic baths or both fermionic and bosonic baths.
This is the subject of current work and
will allow the study of large classes of impurity models,
e.g., so-called Bose Kondo \cite{bosekondo} and Bose-Fermi Kondo
models \cite{bfk}.
These models are known to display intermediate-coupling
fixed points associated with universal local-moment fluctuations.
The Bose-Fermi Kondo model arises in the context of
extended dynamical mean-field theory (EDMFT) \cite{edmft}, where a lattice model
is mapped onto an impurity model with a fermionic bath (representing
conduction electrons) and a bosonic bath (representing bulk spin fluctuations).
The quantum phase transition appearing in the Bose-Fermi Kondo model has
been proposed to describe local quantum critical behavior in EDMFT,
which may be relevant to the physics of certain heavy-fermion quantum
phase transitions.
However, a full numerical solution of the EDMFT equations at $T=0$ has not
been presented to date, due to the lack of suitable impurity solvers.
A version of the bosonic NRG may help to overcome this difficulty.

Other applications of the bosonic NRG can likely be found in the
rapidly developing field of ultracold bosonic gases, where indeed
various realizations of spin-boson physics have been proposed \cite{sbgases}.

Further, the physics of decoherence of qubits naturally leads to variants
of the spin-boson model. Interestingly, the description of $1/f$ noise in
electrical circuits leads to sub-Ohmic damping with $s=0$ (at least over
a certain range of energies).
In this sub-Ohmic parameter regime, the bosonic NRG
is one of the few methods which can give reliable answers, including,
e.g., the existence of a quantum phase transition for $0<s<1$ -- note that
this transition does not appear in the popular non-interacting blip (NIBA)
approximation \cite{Weiss}.
Other modifications of standard spin-boson physics include the influence
of localized modes which interact with the qubit of interest -- those modes
can be represented by a discrete spin system, leading to so-called central
spin models \cite{centralspin}.
Usually, such systems map onto spin-boson models with a spectral density
consisting of a continuous (e.g.~Ohmic) background and sharp peaks at certain
frequencies\cite{structbath}; these models can be easily studied using NRG.

\begin{acknowledgments}
We thank
 M. Garst,
 T.-H. Gimm,
 H. Horner,
 E. Jeckelmann,
 S. Kehrein,
 Th. Pruschke,
 A. Rosch,
 P. W\"olfle,
 and W. Zwerger
for discussions.
This research was supported by the DFG through SFB 484 (RB, HJL) and
the Center for Functional Nanostructures Karlsruhe (MV), and by
the Alexander von Humboldt foundation (NHT).
\end{acknowledgments}

\appendix
\section{Calculation of the Parameters of the Semi-Infinite Chain}
\label{app:A}
In this appendix, we describe the orthogonal transformation
from the star-Hamiltonian (\ref{eq:hstar}) to the
semi-infinite chain form [the chain-Hamiltonian (\ref{eq:hchain})]
and present equations relating
the parameters of the two Hamiltonians.

We start from the star-Hamiltonian (\ref{eq:hstar}):
\begin{equation}
H_{\rm s} = H_{\rm loc}+ \sum_{n=0}^{\infty} {\xi_n a^{\dagger}_n a_n}
           + \frac{\sigma_z}{2 \sqrt{\pi}} \sum_{n=0}^{\infty} {\gamma_n \left( a_n + a^{\dagger}_n \right) }  \,.              \label{B1}
\end{equation}
Our goal is to transform it to a semi-infinite chain
eq.~(\ref{eq:hchain}):
\begin{eqnarray}
H_{\rm c} &=&  H_{\rm loc}+ \sqrt{\frac{\eta_0}{\pi}} \frac{\sigma_z}{2} \left(b_0+ b^{\dagger}_0 \right)
 \nonumber\\
          &+&  \sum_{n=0}^{\infty} \left[ \epsilon_n b^{\dagger}_n b_n + t_n \left( b^{\dagger}_n b_{n+1}
                   +b^{\dagger}_{n+1}b_n \right) \right] \,.   \label{B2}
\end{eqnarray}
Here the main difference to the fermionic
case (such as the Kondo model studied in Ref.~\onlinecite{Wil75}) is that
the bosonic spectral function $J(\omega)$ is restricted to positive
frequencies only. This  asymmetry  of
$J(\omega)$ influences the structure of the semi-infinite chain Hamiltonian
(additional on-site energies $\epsilon_n$ appear in eq.~(\ref{B2}) which
are not present for particle-hole symmetry in the fermionic case).

In the following we define a real orthogonal transformation $U$,
\begin{equation}
     b_n=\sum_{m=0}^{\infty} U_{nm} a_m \, ,   \label{B3}
 \end{equation}
with $U^T U=U U^T =1$, $U^*=U$, so that the inverse transformation
reads
\begin{equation}
     a_n=\sum_{m=0}^{\infty} U_{mn} b_m \, .   \label{B3prime}
 \end{equation}
 Comparing the coupling terms between spin and bosons in
$H_{\rm s}$ and $H_{\rm c}$ gives
\begin{equation}
   b_0=\frac{1}{\sqrt{\eta_0}} \sum_{n=0}^{\infty} \gamma_n a_n \, ,   \label{B4}
\end{equation}
so that
\begin{equation}
        U_{0n}=\frac{\gamma_n}{\sqrt{\eta_0}}        \,. \label{B5}
\end{equation}
The bosonic commutation relation $[b_0, b^{\dagger}_0]=1$
applied to eq.~(\ref{B4}) gives
\begin{equation}
    \eta_0 = \sum_{n=0}^{\infty} \gamma_n^{2}
       = \int_{0}^{\omega_c}{J(x)} \, {\rm d}x  \, .                  \label{B6}
\end{equation}
We are left with the equivalence of the free bosonic part in
$H_{\rm s}$ and $H_{\rm c}$:
\begin{equation}
 \sum_{n=0}^{\infty} {\xi_n a^{\dagger}_n a_n} =
 \sum_{n=0}^{\infty} \left[ \epsilon_n b^{\dagger}_n b_n +
       t_n \left( b^{\dagger}_n b_{n+1}
                   +b^{\dagger}_{n+1}b_n \right) \right] \,.
\label{Bfree}
\end{equation}
To obtain the recursion relations for $\epsilon_n$ and $t_n$, we first
put eq.~(\ref{B3prime}) into the left-hand side of eq.~(\ref{Bfree})
(for the annihilation operators only). We then sort the resulting
equation for  the operators $b_m$. Comparing the prefactors of the
terms containing $b_m$ we obtain for the operator $b_0$:
\begin{equation}
     \sum_{n=0}^{\infty} \xi_n a^{\dagger}_{n} U_{0n}=
      \epsilon_0 b^{\dagger}_0 +t_0 b^{\dagger}_1 \, , \label{B7}
\end{equation}
and for $b_m$ with $m>0$:
\begin{equation}
   \sum_{n=0}^{\infty} \xi_n a^{\dagger}_{n} U_{mn}=
\epsilon_m b^{\dagger}_m +t_m b^{\dagger}_{m+1}
                  +t_{m-1} b^{\dagger}_{m-1} \,. \label{B12}
\end{equation}
The expression for $\epsilon_0$ can be obtained from taking
the commutator between $b_0$ and eq.~(\ref{B7}):
 \begin{equation}
     \epsilon_0 = \sum_{n=0}^{\infty} {\xi_n U_{0n}^2} \,,   \label{B8}
 \end{equation}
From eq.~(\ref{B7}), we also obtain
\begin{equation}
     t_0 b^{\dagger}_1=\sum_{n=0}^{\infty}
\left( \xi_n -\epsilon_0 \right) U_{0n} a^{\dagger}_{n} \, , \label{B9}
\end{equation}
which gives immediately
\begin{equation}
     U_{1n} =\frac{1}{t_0}\left( \xi_n - \epsilon_0\right) U_{0n}
   \,.   \label{B10}
\end{equation}
The value of $t_0$ can be calculated by taking the commutator
with the corresponding adjoint operator on both sides of eq.~(\ref{B9}).
This results in
 \begin{equation}
     t_0 = \frac{1}{\sqrt{\eta_0}} \left[ \sum_{n=0}^{\infty}  \left( \xi_n -\epsilon_0 \right)^2 \gamma_{n}^{2}  \right]^{\frac{1}{2}} \, . \label{B11}
 \end{equation}
Equations (\ref{B5}), (\ref{B8}), (\ref{B10}), and (\ref{B11}) initialize the recursion relations for the
calculation of $\epsilon_m$, $t_m$ and $U_{mn}$. These recursion
relations can be obtained by starting with eq.~(\ref{B12}) and proceeding
in a similar way as above.

The commutator beween $b_m$ and eq.~(\ref{B12}) gives
\begin{equation}
    \epsilon_m = \sum_{n=0}^{\infty} \xi_n U_{mn}^2  \,.  \label{B13}
\end{equation}
From eq.~(\ref{B12}) we also find
\begin{equation}
     t_m b^{\dagger}_{m+1}=\sum_{n=0}^{\infty}
        \left( \xi_n U_{mn}-\epsilon_m U_{mn}
              -t_{m-1}U_{m-1 n} \right) a^{\dagger}_{n} \, . \label{B14}
\end{equation}
From this equation, we obtain the expression for $U_{m+1 n}$:
\begin{equation}
     U_{m+1 n} = \frac{1}{t_m} \left[ \left( \xi_n -
\epsilon_m \right) U_{mn} - t_{m-1} U_{m-1 n} \right]    \, .   \label{B15}
\end{equation}
The values of $t_m$ can be calculated by taking the commutator
with the corresponding adjoint operator on both sides of eq.~(\ref{B14}).
This results in
\begin{equation}
     t_m = \left[ \sum_{n=0}^{\infty} \left[  \left( \xi_n -\epsilon_m \right) U_{mn} -t_{m-1}U_{m-1 n}  \right]^2 \right]^{\frac{1}{2}} \,. \label{B16}
 \end{equation}
Equations (\ref{B13}), (\ref{B15}), and (\ref{B16}) complete the recursion relations for
the calculation of the parameters of the chain Hamiltonian
(\ref{eq:hchain}).

Despite the simple structure of the input spectral function,
$J(\omega)=2\pi\alpha\omega^s$ ($s\ge 0$), we did not succeed in solving the
recursion relations analytically (this is in fact possible for the
particle-hole symmetric soft-gap Anderson model, where the hybridization
function vanishes at the Fermi level as
$\Delta(\omega)=\Delta_0 \vert\omega\vert^r$, see Ref.~\onlinecite{BPH};
due to the particle-hole symmetry, the $\epsilon_n$ vanish and the
recursion relations have a much simpler structure). Therefore the
recursion relations have to be iterated numerically, in a similar way as
for the fermionic case. Note that the derivation of the chain-Hamiltonian
in the asymmetric {\em fermionic} case, where $\epsilon_n\ne 0$, is
very similar to the bosonic case described above. The only differences
are the structure of the coupling between impurity and the bath,
and the fact that all
commutators have to be replaced by anticommutators. For a recent
application of the NRG to a fermionic model with an asymmetric hybridization
function, see Ref.~\onlinecite{Martinek}.

The resulting parameters of the chain-Hamiltonian, $\epsilon_n$ and $t_n$,
both fall off as $\Lambda^{-n}$; in contrast to the fermionic
case where $t_n\propto \Lambda^{-n/2}$. For large $n$, the ratio
$t_n/\epsilon_n$ approaches an $s$-dependent value.

\section{Optimal Bosonic Basis in the Star-NRG}
\label{app:B}
In this appendix, we present details of how we implement the optimal basis for
the bosons
in the star-NRG to overcome the problem of the boson number divergence
when the flow is to the localized fixed point (see the discussions
in Secs.~\ref{sec:oblfp} and \ref{sec:3D}).

In each step of the star-NRG, a new bosonic degree of freedom is
added to the Hamiltonian. The renormalization group
transformation is given by eq.~(\ref{eq:rg-Hs}):

\begin{eqnarray}
 & & H_{N+1,\rm s} = \Lambda H_{N,\rm s}\nonumber \\ &+&
         \Lambda^{N+1} \left[ \xi_{N+1}a_{N+1}^{\dagger}a_{N+1}
  + \frac{\sigma_z}{2\sqrt{\pi}}
      \gamma_{N+1} \left(a_{N+1}+a_{N+1}^{\dagger} \right)
\right] \nonumber \\
\end{eqnarray}

The problem discussed in Sec.~\ref{sec:oblfp} is that upon
approaching the localized
fixed point in the sub-Ohmic case,
the displacements $\theta_N$ for the bosonic site $N$
increase exponentially with $N$, see
eq.~(\ref{eq:shift-star}).
The displacements can be taken into account by constructing an
appropriate basis $\vert s(N+1) \rangle$
for the new bosonic degree of freedom.

To construct this basis, we start with
a simplified Hamiltonian of the form:
\begin{equation}
\bar{H} = a^{\dagger}a +
\theta \sigma_z \left( a^{\dagger} + a \right)
\,\, , \label{A2}
\end{equation}
and proceed as follows: in the first step, we set up an optimized basis
for $\bar{H}$ (optimized in the sense that the lowest lying eigenstates
of $\bar{H}$ are described with only a  small number of basis states).
Then we use this basis, denoted as $\vert s(N+1) \rangle_\theta$,
for the actual NRG iteration, and finally, we describe a
self-consistent procedure
to determine the parameter $\theta$.

Let us first discuss how to construct the optimized basis for $\bar{H}$.
Consider the following operators:
\begin{equation}
 H_{\pm\theta} = a^{\dagger}a \pm \theta \left( a^{\dagger} +a \right) \ .
\label{A3}
\end{equation}
The eigenstates of $H_{\pm\theta}$ are denoted as $ |m \rangle_{\pm\theta}$
($m=0,1,...$).
We obtain
\begin{equation}
   H_{\pm \theta} |m \rangle_{\pm \theta} = \left(m-\theta^2 \right) |m \rangle_{\pm \theta} \, , \label{A5}
\end{equation}
and
\begin{equation}
   |m \rangle_{\pm \theta} = e^{\mp \theta \left( a^{\dagger}-a \right)}
   |m \rangle \, ,   \label{A6}
\end{equation}
with $|m \rangle$ the eigenstates of $a^{\dagger} a$.
The basis states should describe the
$+\theta$ and $-\theta$ displacements on an equal footing; therefore
we proceed with symmetrized eigenstates $|m\rangle_{\rm e/o}$ constructed
in the following way:
\begin{eqnarray}
|m\rangle_{\rm e} = c_{{\rm e},m}
\left[ |m\rangle_{\theta} +  (-1)^{m} |m\rangle_{-\theta} \right]
\nonumber \\
|m\rangle_{\rm o} = c_{{\rm o},m} \left[ |m\rangle_{\theta} - (-1)^{m} |m\rangle_{-\theta} \right]
\nonumber \\
\,\,\,\,  m=0,1,..., \frac{N_{\rm b}}{2}-1  \ ,    \label{A7}
\end{eqnarray}
with normalization constants $c_{{\rm e/o},m}$. Note that here we have
to choose an  even number $N_{\rm b}$.
The even  and odd parity states
are orthogonal to each other,
$_{\rm e}\langle n\vert m \rangle_{\rm o}=0$,
 whereas
states with the same parity are not necessarily orthogonal.
An orthogonalization procedure for both even and odd parity states then
gives the final set of basis states:
\begin{eqnarray}
      |\bar{0} \rangle_{\rm e} &=& |0 \rangle_{\rm e}
\nonumber \\
      |\bar{1} \rangle_{\rm e} &=& C_{{\rm e},1}\left\{ |1\rangle_{\rm e} -
             \!  \phantom{\rangle}_{\rm e}\langle \bar{0} \vert
1 \rangle_{\rm e}  |\bar{0} \rangle_{\rm e} \right \}
\nonumber \\
       |\bar{2} \rangle_{\rm e} &=& C_{{\rm e},2} \left\{ |2 \rangle_{\rm e} -
   \!  \phantom{\rangle}_{\rm e}\langle \bar{1}   |  2\rangle_{\rm e}
            |\bar{1} \rangle_{\rm e} -
           \!  \phantom{\rangle}_{\rm e} \langle  \bar{0} | 2\rangle_{\rm e}
 |\bar{0} \rangle_{\rm e} \right\}
\nonumber \\
      \ldots       \ ,          \label{A8}
\end{eqnarray}
with normalization constants $C_{{\rm e/o},m}$.
The same orthogonalization is performed for the odd parity states.
In this way, we obtain $N_{\rm b}$ orthogonal states,
characterized by the parameter $\theta$, which form
the basis $\vert s(N+1) \rangle_\theta$
for the diagonalization of $ H_{N+1,\rm s}$:
\begin{equation}
   \vert s(N+1) \rangle_\theta = \left\{
          |\bar{0} \rangle_{\rm e},  |\bar{1} \rangle_{\rm e},
          \ldots 
          ,
          |\bar{0} \rangle_{\rm o},  |\bar{1} \rangle_{\rm o},
          \ldots 
          \right\} \ .
\end{equation}
The calculation of the matrix elements
$H_{N+1,{\rm s}} (rs,r^\prime s^\prime)$
(see eqs.~(\ref{eq:threeparts}-\ref{eq:H3s}))
involves matrix elements of the form
\begin{equation}
 \phantom{\rangle}_\theta \langle s(N+1) \vert a_{N+1} +  a_{N+1}^{\dagger} \vert
        s^\prime(N+1) \rangle_\theta \ ,
\end{equation}
To evaluate these matrix elements [and the scalar products in
eq.~(\ref{A8})] we have to express the states
$|\bar{m} \rangle_{\rm e/o}$ in terms of the eigenstates
$\vert n \rangle$ of $a_{N+1}^{\dagger} a_{N+1}$.
This can be performed using the following recursion relations for
$\langle n |m \rangle_{\theta}$:
\begin{eqnarray}
      \langle n |m+1 \rangle_{\theta} &=& \frac{\theta}{\sqrt{m+1}} \langle n | m \rangle_{\theta}
         + \frac{\sqrt{n}}{\sqrt{m+1}} \langle n-1 | m \rangle_{\theta} \, ,
\nonumber \\
      \langle n | 0 \rangle_{\theta} &=& \frac{(-\theta)^{n}}{\sqrt{n!}}e^{-\frac{1}{2}\theta^2} \, ,
\nonumber \\
      \langle 0|m \rangle_{\theta} &=& \frac{\theta^{m}}{\sqrt{m!}}e^{-\frac{1}{2}\theta^2} \, \, , \label{A9}
\end{eqnarray}
(and for $\langle n |m \rangle_{-\theta}$ by replacing $\theta$
by  $-\theta$). The summation over $n$ in the calculation of matrix
elements and scalar products has to be performed numerically which limits
the number of states $|n \rangle$ to some finite, although very large,
value $L$ (values up to $L\approx10^7$ can be used). To construct
an optimized basis for the displacements $\theta$, $L$ should be
large enough (at least of the order of $\theta^2$) to include
a sufficient number of states $|n \rangle$ in the calculation.

For the special case of $\Delta=0$, the parameter $\theta$ for the
construction of the basis $\vert s(N+1) \rangle_\theta$ is exactly
known (see Sec.~\ref{sec:oblfp}). This is different in the general
case of finite $\Delta$ where we have to find a scheme to determine
the optimal value $\theta^\ast$. The general strategy to find
this optimal value has been discussed in Sec.~\ref{sec:genstrat}.
For the actual numerical calculation it turns out that the following
self-consistent scheme is much more efficient.

For the ground state $|g \rangle$
of $\bar{H}$ (\ref{A2}) the expectation value
$\langle g | a^{\dagger} a |g \rangle$ is equal to $\theta^2$. We use
this relation to determine the $\theta$ used for the NRG calculation:
\begin{equation}
   \theta =\sqrt{\phantom{\rangle}_{N+1}
\langle g | a^{\dagger}_{N+1} a_{N+1} |g \rangle_{N+1}} \ ,   \label{A10}
\end{equation}
where $|g \rangle_{N+1}$ is the ground state of
$H_{N+1,\rm s}$ which has been obtained from diagonalizing
the matrix $H_{N+1,{\rm s}} (rs,r^\prime s^\prime)$
using the basis $\vert s(N+1) \rangle_\theta$ characterized
by the parameter $\theta$. In other words, eq.~(\ref{A10})
defines a self-consistent scheme to calculate
$\theta$ for each NRG step.

The converged value $\theta^\ast$ gives the optimal basis for
adding the site $N+1$ in the NRG iteration. It corresponds to
the value $\theta^\ast$ which characterizes the minimum
of the energy levels in Fig.~\ref{fig:Evstheta}. The energy
levels calculated in this way show a much weaker dependence on
$N_{\rm b}$ which leads, for example, to the rapid convergence
of $n_{\rm b}$ with increasing $N_{\rm b}$ as
shown in Fig.~\ref{fig:n_divergence}.

After the diagonalization of $H_{N+1,\rm s}$ with the optimized
basis $\vert s(N+1) \rangle_{\theta^\ast}$, we can continue the NRG
iteration by adding the site $N+2$.


\addcontentsline{toc}{section}{Bibliography}
\begin{thebibliography}{99}

\bibitem{Wil75}
  K. G. Wilson,  Rev. Mod. Phys. {\bf 47}, 773 (1975).
\bibitem{Kri80}
  H. R. Krishna-murthy, J. W. Wilkins, and K. G. Wilson,
  Phys. Rev. B {\bf 21}, 1003 (1980); {\it ibid.} {\bf 21}, 1044 (1980).
\bibitem{Hewson} A. C. Hewson,
  {\it The Kondo Problem to Heavy Fermions} (Cambridge Univ. Press,
  Cambridge 1993).
\bibitem{Costi99a} T. A. Costi, in {\it Density-Matrix Renormalization -
    A New Numerical Method in Physics}, Eds. I.~Peschel {\em et al.}
    (Springer 1999).
\bibitem{RBAdvPhys}
  R. Bulla, Adv. Solid State Phys. {\bf 40}, 169 (2000).
\bibitem{multinrg}
  O. Sakai and Y. Shimizu,
  J. Phys. Soc. Jpn. {\bf 61}, 2333 (1992);
  {\it ibid.} {\bf 61}, 2348 (1992);
  O.~Sakai, Y. Shimizu, and N. Kaneko,
  Physica B {\bf 186}-{\bf 188}, 323 (1993).
\bibitem{twochnrg}
  K. Ingersent and B. A. Jones, Physica B {\bf 199-200}, 402 (1994).
\bibitem{holsteinnrg}
  A.~C. Hewson and D. Meyer, 
  J. Phys. Condens. Matter {\bf 14}, 427 (2002);
  D. Meyer, A.~C. Hewson, and R. Bulla, 
  Phys. Rev. Lett. {\bf 89}, 196401 (2002).
\bibitem{MV} W. Metzner and D. Vollhardt,
   Phys. Rev. Lett. {\bf 62}, 324 (1989).
\bibitem{Geo96}  A. Georges, G. Kotliar, W. Krauth, and
  M. J. Rozenberg, Rev.  Mod. Phys. {\bf 68}, 13 (1996).
\bibitem{dmftnrg}
  R.~Bulla, \prl {\bf 83}, 136 (1999).
\bibitem{BCV}
  R. Bulla, T. A. Costi, and D. Vollhardt,
  \prb {\bf 64}, 045103 (2001).
\bibitem{BTV} R. Bulla, N. Tong, and M. Vojta,
              Phys. Rev. Lett. {\bf 91}, 170601 (2003).
\bibitem{Leggett}
   A.~J. Leggett, S. Chakravarty, A. T. Dorsey, M. P. A. Fisher,
  A. Garg, and W. Zwerger,
  Rev. Mod. Phys. {\bf 59}, 1 (1987).
\bibitem{Weiss} U. Weiss, {\it Quantum dissipative systems}, 2nd ed.
    (World Scientific, Singapore, 1999).
\bibitem{structbath}
   S. Kleff, S. Kehrein, and J. von Delft, Physica E {\bf 18}, 343 (2003);
   F.~K. Wilhelm, S. Kleff, and J. von Delft, Chem. Phys. {\bf 296}, 345
   (2004).
\bibitem{centralspin}
  N. Prokov\'ev and P. C. E. Stamp, Rep. Prog. Phys. {\bf 63}, 669 (2000);
  P. C. E. Stamp and I. S. Tupitsyn, 
  Chem. Phys. {\bf 296}, 281 (2004).
\bibitem{Cos03}
  T. A. Costi and R. H. McKenzie, Phys. Rev. A {\bf 68}, 034301 (2003).
\bibitem{dima}
  D. V. Khveshchenko, Phys. Rev. B {\bf 69}, 153311 (2004).
\bibitem{wilhelm}
  M. Thorwart and P. H\"anggi, Phys. Rev. A {\bf 65}, 012309 (2002);
  M. J. Storcz and F. K. Wilhelm, {\em ibid.} {\bf 67}, 042319 (2003).

\bibitem{bosekondo}
  S. Sachdev, C. Buragohain, and M. Vojta, Science {\bf 286}, 2479 (1999);
  A. H. Castro Neto, E. Novais, L. Borda, G. Zar\'and, and I. Affleck,
  Phys. Rev. Lett. {\bf 91}, 096401 (2003).

\bibitem{Garg} A. Garg, J.~N. Onuchic, and V. Ambegaokar,
    J. Chem. Phys. {\bf 83}, 4491 (1985).
\bibitem{MuehlbacherEgger}
   L. M\"uhlbacher and R. Egger,
   J. Chem. Phys. {\bf 118}, 179 (2003);
   Chem. Phys. {\bf 296}, 193 (2004).

\bibitem{qmcsb}
  R. Egger and C. H. Mak, Phys. Rev. B {\bf 50}, 15210 (1994);
  K. V\"olker, Phys. Rev. B {\bf 58}, 1862 (1998).

\bibitem{nishiyama}
  Y. Nishiyama, Eur. Phys. J. B {\bf 12}, 547 (1999).

\bibitem{Costi98}
   T. A. Costi,
   Phys. Rev. Lett.  {\bf 80}, 1038 (1998).

\bibitem{MVLF}
   Both the sub-Ohmic spin-boson model and the pseudogap Kondo model
   display non-trivial quantum phase transitions, which are, however,
   in different universality classes, see:
   M. Vojta and L. Fritz, Phys. Rev. B {\bf 70}, 094502 (2004);
   L. Fritz and M. Vojta, cond-mat/0408543.

\bibitem{KM} 
   S. Kehrein and A. Mielke, Phys. Lett. A {\bf 219}, 313 (1996).

\bibitem{spohn} 
   H. Spohn and R. D\"umcke, J. Stat. Phys. {\bf 41}, 389 (1985).

\bibitem{koster}
   J. M. Kosterlitz, \prl {\bf 37}, 1577 (1976).

\bibitem{VTB}
   Results for the critical exponents in the sub-Ohmic case
   are shown in:
   M. Vojta, N.-H. Tong, and R. Bulla,
   cond-mat/0410132. 

\bibitem{BPH} R. Bulla, Th. Pruschke, and A. C. Hewson,
   J. Phys. Cond. Matter {\bf 9}, 10463 (1997).
\bibitem{footalex}
  For Jahn-Teller systems, the mapping to a bosonic chain
  was also considered in:
  S. N. Evangelou and A. C. Hewson, J. Phys. C {\bf 15}, 7073 (1982).
\bibitem{Jeckelmann} A similar problem appears in DMRG calculations
   for the one-dimensional Holstein model, see:
   C. Zhang, E. Jeckelmann, and S. R. White,
    Phys. Rev. Lett. {\bf 80}, 2661 (1998).
\bibitem{AYH}
   P. W. Anderson, G. Yuval, and  D. R. Hamann,
   \prb {\bf 1}, 4464 (1970).
\bibitem{BGLP}
   R. Bulla, M. T. Glossop, D. E. Logan, and Th. Pruschke.
   J. Phys.: Condens. Matter {\bf 12}, 4899 (2000).
\bibitem{Oli94}
   W. C. Oliveira and L. N. Oliveira,
   Phys. Rev. B {\bf 49}, 11986 (1994).

\bibitem{expnote}
   The relation
   $T^* \propto \Delta^{1/(\alpha_c-\alpha)}$ 
   (see Ref.~\onlinecite{Leggett}) for the
   crossover scale in the Ohmic spin-boson model, obtained from various
   perturbative approaches, breaks down very close to $\alpha=1$, where
   it has to be replaced by
   $T^* \propto e^{{\rm const}/\sqrt{\alpha_c-\alpha}}$, see:
   J.~M.~Kosterlitz, J. Phys. C {\bf 7}, 1046 (1974).
   Our numerical data in Fig.~\ref{fig:3_8} can be equivalently fitted
   with both forms, and the resulting values of $\alpha_{\rm c}$
   differ by less than 2\%.

\bibitem{CZ}
   T. A. Costi and G. Zar\'and,
   Phys. Rev. B {\bf 59}, 12398 (1999).
\bibitem{Goehrlich88} R. G\"ohrlich and U. Weiss,
           Phys. Rev. B {\bf 38}, 5245 (1998).
\bibitem{Sassetti90} M. Sassetti and U. Weiss,
           Phys. Rev. Lett. {\bf 65}, 2262 (1990).
\bibitem{BH} R. Bulla and A. C. Hewson,
          Z. Phys. B {\bf 104}, 333 (1997).
\bibitem{Fro86}
     H. O. Frota and L. N. Oliveira,
      Phys. Rev. B {\bf 33}, 7871 (1986).
\bibitem{Sak89}
  O. Sakai, Y. Shimizu,  and T. Kasuya,
   J. Phys. Soc. Jpn. {\bf 58}, 3666 (1989).
\bibitem{Cos92}
    T. A. Costi and A. C. Hewson,
      Philos. Mag. B {\bf 65}, 1165 (1992).
\bibitem{Cos94}
  T. A. Costi, A. C. Hewson,  and V. Zlati\'c,
  J. Phys.: Condens. Matter {\bf 6}, 2519 (1994).
\bibitem{Costi96}
   T.~A. Costi and C. Kieffer,
   Phys. Rev. Lett. {\bf 76}, 1683 (1996).

\bibitem{delta-peak}
A  $\delta(\omega)$ contribution to dynamic quantities
has been discussed within the NRG for fermionic systems in
S. C. Bradley, R. Bulla, A. C. Hewson, and G.-M. Zhang,
Eur. Phys. J. B {\bf 11}, 535 (1999);
M. Vojta and R. Bulla,
Phys. Rev. B {\bf 65}, 014511 (2001).


\bibitem{bfk}
A.~M.~Sengupta, Phys. Rev. B {\bf 61}, 4041 (2000);
L. Zhu and Q. Si, Phys. Rev. B {\bf 66}, 024426 (2002);
G. Zar\'and and E. Demler, Phys. Rev. B {\bf 66}, 024427 (2002);
M. Kir\'can and M. Vojta, Phys. Rev. B {\bf 69}, 174421 (2004).

\bibitem{edmft}
Q.~Si, S.~Rabello, K.~Ingersent, and J.~L.~Smith, Nature
{\bf 413}, 804 (2001); \prb {\bf 68}, 115103 (2003).

\bibitem{sbgases}
A. Recati, P. O. Fedichev, W. Zwerger, J. von Delft, and P. Zoller,
cond-mat/0212413; cond-mat/0404533.

\bibitem{Martinek} J. Martinek, M. Sindel, L. Borda,
     J. Barnas, R. Bulla, J. K\"onig, G. Sch\"on,
     S. Maekawa, and J. von Delft,
     cond-mat/0406323.


\end{thebibliography}
\end{document}